\documentclass[twocolumn]{aastex631}
\usepackage{CJK}

\newcommand{\kms}{$\rm km\ s^{-1}$}
\newcommand{\hii}{H{\sc ii}}
\newcommand{\msun}{M_\odot}

\accepted{2026 March 16}
\submitjournal{ApJ}
\begin{document}
\begin{CJK*}{UTF8}{gbsn}

\title{Kinematics of Wolf-Rayet Stars in the LMC:  Clues to Subtype Origins}

\author[0009-0006-2593-1161]{Caden Burkhardt}
\affiliation{University of Michigan, Department of Astronomy, 1085 S. University, Ann Arbor, Michigan 48109}

\author[0009-0008-6632-8937]{Fiona Han（韩宇菲）}
\affiliation{University of Michigan, Department of Astronomy, 1085 S. University, Ann Arbor, Michigan 48109}
\affiliation{Department of Astronomy, University of Arizona, Tucson, Arizona 85721}

\author[0000-0002-5808-1320]{M. S. Oey (黄香莉)}
\affiliation{University of Michigan, Department of Astronomy, 1085 S. University, Ann Arbor, Michigan 48109}

\author[0000-0001-6251-5315]{Natalia Ivanova}
\affiliation{Department of Physics, University of Alberta, Edmonton, T6G 2E7, Alberta, Canada}

\author[0000-0002-6718-9472]{Mathieu Renzo}
\affiliation{Department of Astronomy, University of Arizona, Tucson, Arizona 85721}

\begin{abstract}
We measure transverse proper motion velocities of LMC Wolf-Rayet (WR) stars using Gaia DR3 astrometry. The combined velocity distribution of WNh, O If*/WN, and WNL very massive stars ($>100\ M_\odot$; VMS) shows both slow, unejected objects ($v_\perp < 10$ \kms) and stars dominated by fast, runaway velocities ($v_\perp > 24$ \kms).  This supports expectations that VMS ages are comparable to the dynamical ejection timescale ($\sim1.5$ Myr). These kinematics share similarities with those of lower-luminosity, classical WNh, O If*/WN, and WNL stars, as well as the SMC field OB stars, suggesting that dynamical ejections may also dominate these populations. In contrast, both single and binary WNE stars are ejected populations that show single-peaked velocity distributions, suggesting a different ejection mechanism(s). We speculate that single WNE stars might result from explosive mergers onto the shell-burning layer, thereby stripping the H envelope. Binary WC stars appear to be faster (median $v_\perp = 54$ \kms) and have higher luminosities than singles (median $v_\perp = 38$ \kms), suggesting that single WC stars are not descendants of the binaries. Thus, the binaries are probably stripped by mass transfer, while the WC singles likely originate from another process. The high velocities of binary WC stars are consistent with some predictions that lower mass clusters generate fast dynamical ejections. Single WC and WN3/O3 stars have ambiguous kinematics, but both show high $v_\perp$ (median $\sim 38$ \kms), possibly linked to their lower masses. 

\end{abstract}

%% \keywords{\uat{Wolf-Rayet stars}{1806},\uat{Stellar dynamics}{1596},\uat{Stellar evolutionary models}{2046},\uat{Interacting binary stars}{801},\uat{Runaway stars}{1417},\uat{Magellanic Clouds}{990},\uat{Late-type galaxies}{907},\uat{Massive stars}{732},\uat{Stellar populations}{1622},\uat{Stellar mergers}{2157},\uat{Stellar atmospheres}{1584},\uat{Young massive clusters}{2049}}


\section{Introduction} 
\label{sec:intro}
Massive stars, with their high luminosities and strong stellar winds, have a profound effect on the ionization state, kinematics, and morphology of the surrounding interstellar medium (ISM). Their impact is responsible for driving the evolutionary phenomena of their host galaxies and ultimately of the universe.  One of the stellar phases generating the most energetic feedback effects is the Wolf-Rayet (WR) stage, and most of these stars, furthermore, are progenitors of supernovae, neutron stars, black holes, and/or energetic transients. Therefore, clarifying the origin and nature of WR stars is crucial to understanding massive star evolution and its end products.

WR stars are characterized by distinctive and broad emission lines, and their presence is a widely used diagnostic of massive star populations and their ages both locally and in more distant WR galaxies \citep[see, e.g.,][]{Schaerer1999}. They are mostly believed to be the exposed cores of stars, enriched with helium-burning products, whose envelopes have been removed either by self-stripping due to their strong stellar winds \citep[e.g.,][]{Conti1975} or by binary mass transfer \citep[e.g.,][]{Maeder1991}. Some very massive stars (VMS) $> 100\ \msun$ are believed to be H-burning stars that show WR features simply because their large Eddington parameters $\Gamma_e$ drive them towards extreme mass loss rates \citep[e.g.,][]{deKoter1997, Crowther2010, Grafener2011, Brands2022}.
The different WR spectral subtypes derive in part from their progenitor masses and the degree to which the star has been stripped \citep[e.g., review by ][]{Langer2012}; in the classical scenario, the WN, WC, and WO subtypes are believed to represent progressively stripped cores, and they are defined by the relative strengths of observed N, C, and O emission lines. While WR subtypes are thus linked to evolutionary status, we note that the WR phenomenon is fundamentally a signature of the stellar wind \citep[e.g.,][]{Gotberg2023} and thus also depends on the wind strength, mass-loss rate, and opacity. In this work, we assume that stars of similar subtype and luminosity generally share the same origin, but we note that this may be an overgeneralization.

In addition, while stellar mass loss rates decrease with decreasing metallicity, the threshold mass for generating WR features increases, and thus it is somewhat unclear what role metallicity plays in generating WR stars via single vs binary routes \citep{Shenar2020}. \citet{Pauli2022} find that the LMC WR population may be dominated by objects with binary histories. Thus, the multiple possible evolutionary paths to generate WR features \citep[e.g.,][]{Shenar2019a} include:

\begin{enumerate}
    \item Single, very massive ($>$100 $M_\odot$), core H-burning stars \citep[e.g.,][]{Crowther2010}, that could appear as O If*/WN or WNh stars
    \item Traditional, single WR stars self-stripped by their strong stellar winds, which may be more prevalent at high metallicity \citep[][]{Conti1975}
    \item Binary mass donors stripped by companion interaction \citep[e.g.,][]{Paczynski1967, Maeder1991}
    \item Binary mass gainers that are forced into the self-stripping regime by their enhanced mass and/or rotation \citep[e.g.,][]{Cantiello2007, Eldridge2011}
    \item Binary mergers where mass loss and a subsequently less bound envelope enhance self-stripping \citep{Li2024}.  In this work, we also suggest that explosive nucleosynthesis \citep[e.g.,][]{Ivanova2003} could be another possible mechanism for envelope ejection (Section~\ref{sec:merger}).
\end{enumerate}

Stellar kinematics provide an important tool for clarifying the origins of the different WR classes, for example, by comparing the stellar evolution timescales with ejection timescales.  There are two principal mechanisms that have been considered in the literature for accelerating stars \citep{Hoogerwerf2001}, the binary supernova scenario (BSS) and dynamical ejection scenario (DES). Supernova ejections can eject the surviving companions at their pre-explosion orbital velocity \citep{Blaauw1961}, which depends mostly on the pre-supernova orbital evolution regardless of the nature of the compact object formed \citep{Wagg2025b, Renzo2019a}. The dynamical mechanism is also common and is primarily due to binary-binary interactions deep in clusters \citep{Poveda1967}. Stars with spatial velocities $> 30$ \kms\ are conventionally defined to be “runaway” stars \citep[e.g.][]{Blaauw1956, Gies1986, Eldridge2011} with ``walkaway" stars defined to be unbound objects with velocities below this threshold. In general, BSS ejections are on average slower and generally expected to be walkaways \citep[e.g.,][]{Renzo2019a}, with lower-mass stars accelerated to higher velocities; whereas DES ejections are faster, with the highest-mass stars accelerated the most. \citet{Oey2018} and \citet{DorigoJones2020} demonstrated that the effects of these different mechanisms can be differentiated kinematically in a stellar population, and there is some evidence that DES may dominate the observed field massive-star population \citep[e.g.,][]{Carretero-Castrillo2025, Moe2025, Phillips2024, Carretero-Castrillo2023, DorigoJones2020}. Moreover, binaries initially ejected via the DES mechanism can subsequently be further accelerated by the BSS mechanism \citep[two-step ejections;][]{PflammAltenburg2010}.  However, a different acceleration mechanism that deserves further study is acceleration by merger events; we suggest this as a possibility in Section~\ref{sec:merger}.

We use the tools our group developed to study the kinematics of other massive star populations such as OB, OBe, B[e], and luminous blue variables (LBVs) in the SMC and LMC to similarly study WR stars in the LMC.

\section{LMC WR Kinematics}

\subsection{Sample and Measurements}\label{sec:meas}

To obtain the transverse velocities of WR stars in the LMC, we use the Fifth Catalog of LMC Wolf-Rayet Stars \citep{Neugent2018}, which offers a complete population of 156 WR objects, and we cross-match this with proper motion data from the Gaia DR3 release \citep[][]{Gaia2016,Gaia2023}.

To identify our WR stars in the Gaia DR3, we follow the general method outlined in \cite{Oey2018}. We specify a $3\arcsec$ position cross-match between the cataloged WR stars and Gaia DR3. To distinguish our target WR stars from their neighbors in crowded regions, we eliminate all objects whose $V$ magnitudes differ from those of the targets by more than 0.3 mag, using photometric corrections taken from \citet{Riello2021} to convert Gaia $G$-band magnitudes into $V$-band magnitudes in the Johnson-Cousins system. We also eliminate objects with parallax $> 0.25$ mas to remove Galactic foreground stars, and we eliminate all objects with incomplete photometry or incomplete proper motion data. We eliminate all stars with either R.A. or decl. position errors $>$ $3.5\sigma$ from the median errors in both R.A. and decl., which are 0.023 mas for both; and those with either R.A. or decl. proper motion error $>1.5\sigma$ from the median errors of 0.029 and 0.031  mas yr$^{-1}$, respectively. We remove BAT99 101 and BAT99 102 from our sample; these two stars are shown at the same position in the \citet{Neugent2018} catalog,  it is therefore unclear which one is the correct match to the single, corresponding Gaia object. 

Overall, we retain a sample of 119 Gaia objects, which correspond to 121 WR stars, since two systems, BAT99 118 and BAT99 119, are double counted as WR + WR binaries. We refer to this sample as our ``Gaia sample" and the 154 objects or 156 WR stars in the complete catalog of \citet{Neugent2018} as the ``Total" population in what follows.

To calculate the residual transverse velocity $v_\perp$ for each star in our Gaia sample, we follow the general method of \cite{Phillips2024}. We use the DR3 proper motion values obtained from \cite{Gaia2023} for each target star and its local field stars, defined as all stars with $G\leq 18$ within a $3\arcmin$ radius. We adopt a distance of 49.59 kpc to the LMC \citep[][]{Pietrzynski2019}. Our final velocities are obtained by subtracting the median R.A. and decl. field velocities from those of the target star.  Our resulting $v_\perp$ for all the Gaia sample WR stars are given in the Appendix, Table \ref{tab:Full}.  

\begin{longrotatetable}
\begin{deluxetable*}{lccccccccc}
\tablecaption{Kinematic Properties of WR Populations}
\label{tab:subpop}
\tablehead{
\colhead{Population} & 
\colhead{N} & 
\colhead{N} & 
\colhead{median $v_\perp$} & 
\colhead{wt med $v_\perp$} &
\colhead{std err} & 
\colhead{N (30 Dor)} & 
\colhead{N (30 Dor)} &
\colhead{wt med $v_\perp$} & 
\colhead{wt med $v_\perp$} \\
& 
\colhead{Total}& 
\colhead{Gaia}& 
&
&
&
\colhead{Total}& 
\colhead{Gaia}&
\colhead{30 Dor} &
\colhead{Non-30 Dor} 
}
\decimalcolnumbers
\startdata
Total & 156 & 121 & 30 & 28 & 3.0 & 48 & 29 & 21 & 30 \\
Single & 118 & 91 & 31 & 30 & 3.3 & 41 & 23 & 20 & 34 \\
Binary & 38 & 30 & 24 & 23 & 7.1 & 7 & 6 & 37 & 23 \\
Classical WNh, O If*/WN, WNL\tablenotemark{a} & 23 & 21 & 23 & 22& 7.6 & 4 & 3 & 20 & 23 \\
VMS WNh, O If*/WN, WNL\tablenotemark{b}       & 28 & 17 & 37 & 21 & 7.1 & 26 & 15 & 21 & 3.6\tablenotemark{c} \\
WN\tablenotemark{d}        & 120 & 94 & 28 & 23 & 3.3 & 40 & 25 & 20 & 25 \\
\hline
WNh\tablenotemark{e} & 22 & 16 & 22 & 18 & 7.1 & 11 & 5 & 43 & 18 \\
WNh (Classical)\tablenotemark{a}     & 11 & 10 & 20 & 18 & 8.9 & 1 & 0 & $\cdots$ & 18 \\
WNh (VMS)\tablenotemark{b}             & 10 & 5 & 43 & 43 & 14 & 9 & 4 & 52 & 42 \\
\hline
O If*/WN\tablenotemark{e} & 13 & 10 & 34 & 20 & 13 & 11 & 8 & 20 & 48 \\
O If*/WN (Classical)\tablenotemark{a}      & 3 & 3 & 56 & 20 & 20 & 2 & 2 & 20 & 74 \\
O If*/WN (VMS)\tablenotemark{b}             & 8 & 5 & 15 & 8.9 & 8.0 & 8 & 5 & 8.9 & $\cdots$ \\  
\hline
WNL\tablenotemark{d,e} & 23 & 16 & 37 & 23 & 9.0 & 10 & 7 & 21 & 38 \\
WNL (Classical)\tablenotemark{a}      & 9 & 8 & 43 & 38 & 14 & 1 & 1 & 16 & 50 \\
WNL (VMS)\tablenotemark{b}             & 10 & 7 & 37 & 21 & 12 & 9 & 6 & 37 & 3.6 \\
\hline
WNE\tablenotemark{d} & 60 & 51 & 28 & 26 & 4.4 & 8 & 5 & 16 & 28 \\
WNE Single & 39 & 33 & 30 & 30 & 5.8 & 7 & 4 & 16 & 30 \\
WNE Binary & 21 & 18 & 20 & 17 & 6.0 & 1 & 1 & 15 & 23 \\
\hline
WC        & 23 & 16 & 43 & 42 & 10 & 7 & 3 & 38 & 43 \\
WC Single & 13 & 9 & 38 & 38 & 6.3 & 6 & 2 & 21 & 42 \\
WC Binary & 10 & 7 & 54 & 54 & 22 & 1 & 1 & 54 & 53 \\
\hline
WN3/O3 & 10 & 10 & 38 & 39 & 8.3 & 1 & 1 & 17 & 39 \\
WO & 3 & 1 & 57 & 57 & 0.0 & 0 & 0 & $\cdots$ & 57 \\  
\enddata
\tablenotetext{a}{We define classical WNh, O If*/WN, and WNL stars to be those with $L/L_\odot \leq 5.9$}
\tablenotetext{b}{We define VMS WNh, O If*/WN, and WNL stars to be those with $L/L_\odot > 5.9$ }
\tablenotetext{c}{There are only 2 VMS stars outside 30 Dor, a VMS WNL with $v_\perp = 3.6$  \kms\ and VMS WNh with 42 \kms.}
\tablenotetext{d}{Excluding WN3/O3 stars. One WN star in the \cite{Neugent2018} catalog does not have a subtype, so it is not counted as either a WNE or WNL star.}
\tablenotetext{e}{Classical and VMS subgroups are defined by their luminosities, so stars without luminosity estimates are excluded from these two subgroups but are included in the subtype totals.}

\end{deluxetable*}
\end{longrotatetable}

\subsection{Velocity Distributions}

\begin{figure}
    \centering
    \includegraphics[width=1.0\linewidth]{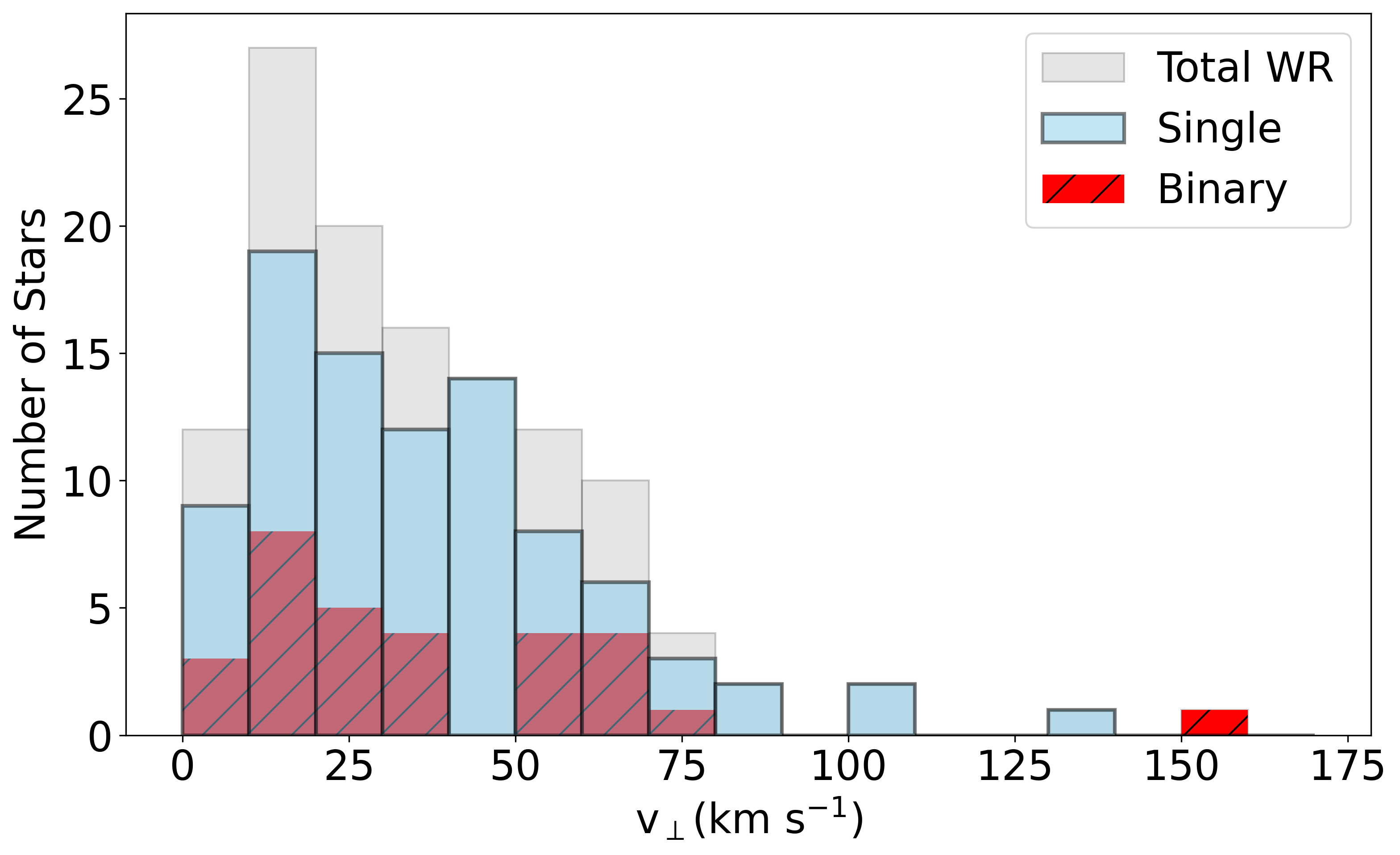}
    \caption{Contributions of single (blue) vs binary (red hatched) WR stars toward to the total WR velocity distribution in the LMC.  These two distributions are overplotted, not stacked.}
    \label{fig:Single vs Binary}
\end{figure}

We compile the median velocities for individual WR subtypes and subgroups in Table \ref{tab:subpop}. The median values given throughout this work correspond to the medians weighted by the inverse of their errors (wt med).  WN stars dominate the LMC WR population, accounting for over three quarters of our Gaia sample. Of these, $54\pm9$\% are WNE stars, which have a median velocity of 26 \kms.  The WNL, WNh, and O If*/WN subtypes all show median velocities around 20 \kms, whereas the WC and WN3/O3 stars show faster median velocities around 40 \kms.  We discuss the kinematics of the individual subtypes below. We do not discuss WO stars, as there is only one in our Gaia sample, which is insufficient for analysis of the subtype as a whole.

Figure \ref{fig:Single vs Binary} shows the total distributions of $v_\perp$ of single and binary WR stars in the LMC. The conventional 3-D runaway velocity of 30 \kms\  corresponds to an average $v_\perp$ of 24 \kms. With this threshold definition, over half of the WR stars in our Gaia sample (68/121 or $56\pm9$\% ) are runaways, with a weighted median $v_\perp$ of 24 \kms. These velocities are consistent with these stars having been ejected from clusters as runaway or walkaway stars. We find that single WR stars have higher median $v_\perp$ (30 \kms) than binary  stars (23 \kms; Figure \ref{fig:Single vs Binary}). We use the binary classifications of \citet{Neugent2018}, who adopt as binaries stars whose spectral types are given as ``+ abs" (Table~\ref{tab:Full}); thus, when discussing ``binaries" in what follows, by default we consider only pre-SN binaries that are non-compact objects. However, this may underestimate the total number of binaries, missing those with smaller companions hidden by their more luminous counterparts. As such, our binary sample may be incomplete. In what follows, missing binary identifications generally does not change conclusions we make for stars that are dynamically ejected, although dynamically ejected stars with binary companions should have lower velocities than single stars \citep{Oh2016}. However, in some discussions below, we suggest that single vs binary populations may show distinct kinematics, and therefore, possibly distinct origins. Misclassifications in single vs binary stars could affect these interpretations.

\subsection{30 Doradus}\label{subsec:30Dor}

\begin{figure*}
    \centering
    \includegraphics[height=7in]{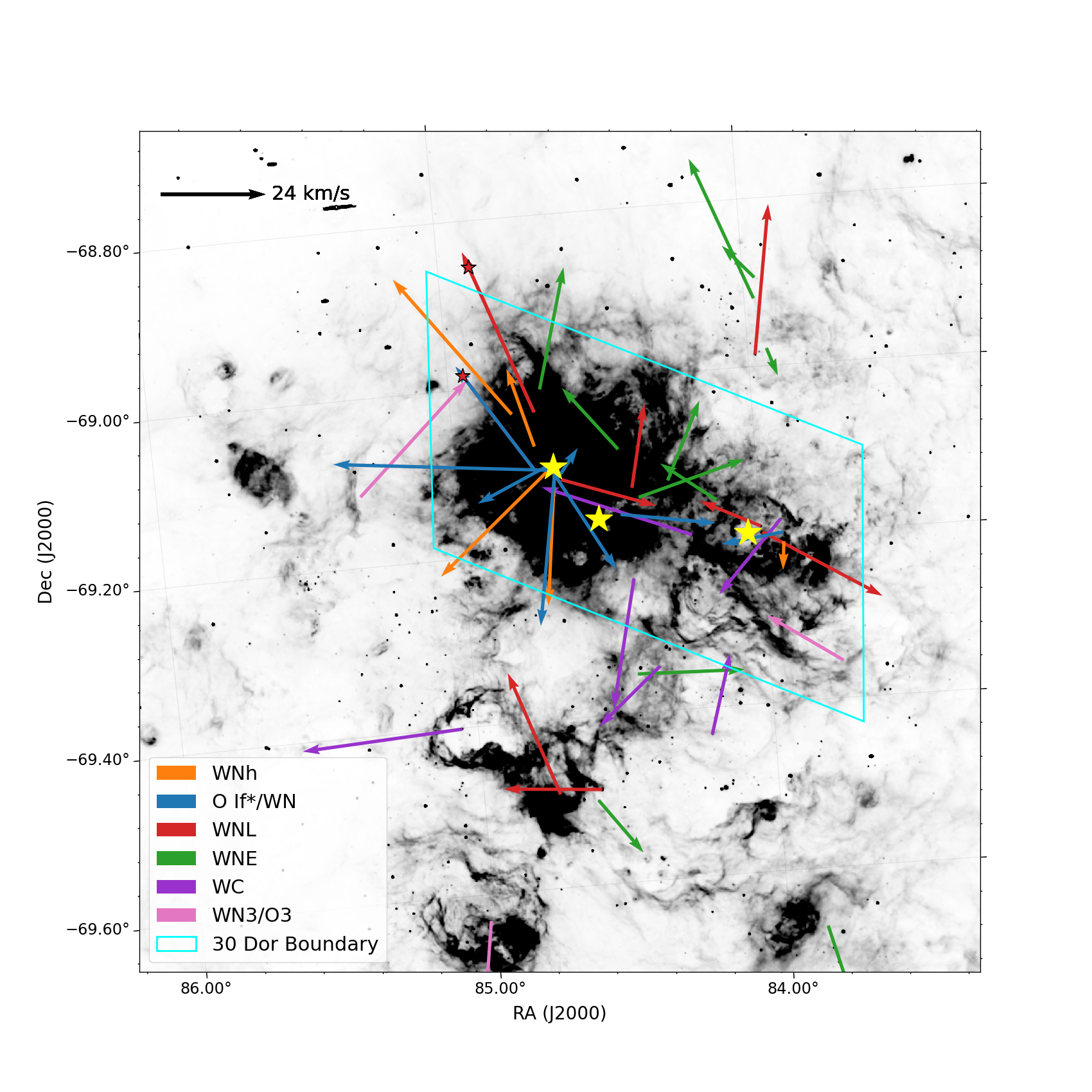}
    \caption{Proper motion vectors for all WR subtypes in the 30 Dor region
    superposed on H$\alpha$ image of the LMC from \citet{Smith2005}, color-coded as shown. Vector lengths are scaled as $v_\perp^{0.5}$. The bases of the vectors are plotted at the WR-star positions. WR+WR binary systems are indicated with stars on the vector arrow heads, color-coded according to the companion subtype, which is WNL in both cases.  
    The three large star symbols indicate the locations of the dominant star-forming regions (east to west): 30 Dor A, B, and C (LH 100, 99, and 90, respectively) \citep{LeMarne1968,Lucke1970}.  The cyan boundary indicates the general region within which stars are considered to be associated with 30 Dor by \citet{Hung2021}. }
    \label{fig:vectors_30Dor}
\end{figure*}

We give special consideration to the 30 Doradus region. This giant star-forming complex is dominated by R136, one of the youngest, densest, and most massive clusters in the Local Group. Mass segregation in R136 causes the most massive stars to fall into the core center, where gravitational interactions can subsequently eject these VMS via the DES mechanism. R136 is estimated to be $\sim 1-2$ Myr old \citep{Crowther2016}, which is comparable to the predicted dynamical ejection timescale for O-type stars in dense, massive clusters \citep[e.g.,][]{Oh2015,Oh2016}, and too young to have significant contributions from BSS survivors.  Thus, the fraction of massive stars ejected from this object  can probe its dynamical ejection timescale. DES ejections from this system have been individually identified and studied \citep[e.g.,][]{PortegiesZwart2025, Renzo2019b, 
% Shenar2019a, 
Lennon2018}, and \citet{Stoop2024} suggest that up to 1/3 of the massive stars from this $\sim 1-2$ Myr-old super star cluster are already ejected. Thus, this rich object introduces unique kinematic signatures and an opportunity to quantitatively study the DES. 

\citet{Hung2021} identify WR stars in the 30 Dor region to be those encompassed by the Tarantula Nebula, identified as DEM L 263 in the \cite{Davies1976} catalog. This area includes the subregions 30 Dor A, B, and C \citep{LeMarne1968}, which correspond to OB associations LH 100, 99, and 90 in \citet{Lucke1970}.  R136 is found within 30 Dor A (LH 100), and stars within a $20\arcsec$ radius of BAT99 108 (R136a1) are defined to be within this cluster, following \citet{Doran2013}. Figure \ref{fig:vectors_30Dor} shows all of our 30 Dor, Gaia-sample stars based on the \citet{Hung2021} definition.  They constitute 29 out of 121, or 24 $\pm$ 5\% of our sample, and they are identified with asterisks in Table \ref{tab:Full}.

The high crowding in R136 leads to incomplete sampling and/or high proper motion errors for stars in the core of this object. Of the 35 out of 156 WR stars with no usable Gaia data, over half (19) are from the 30 Dor region. Of these, 8 stars are in R136 and are likely omitted due to crowding. We therefore caution that our 30 Dor subsample is biased against stars found in the cluster cores; these regions are likely dominated by unejected stars, which are expected to have lower velocities \citep[e.g.,][]{Sana2022}. 

\section{WNh, O If*/WN, and WNL Stars}\label{sec:WNLtypes}

\begin{figure*}[t!]
    \centering
    \includegraphics[height=7in]{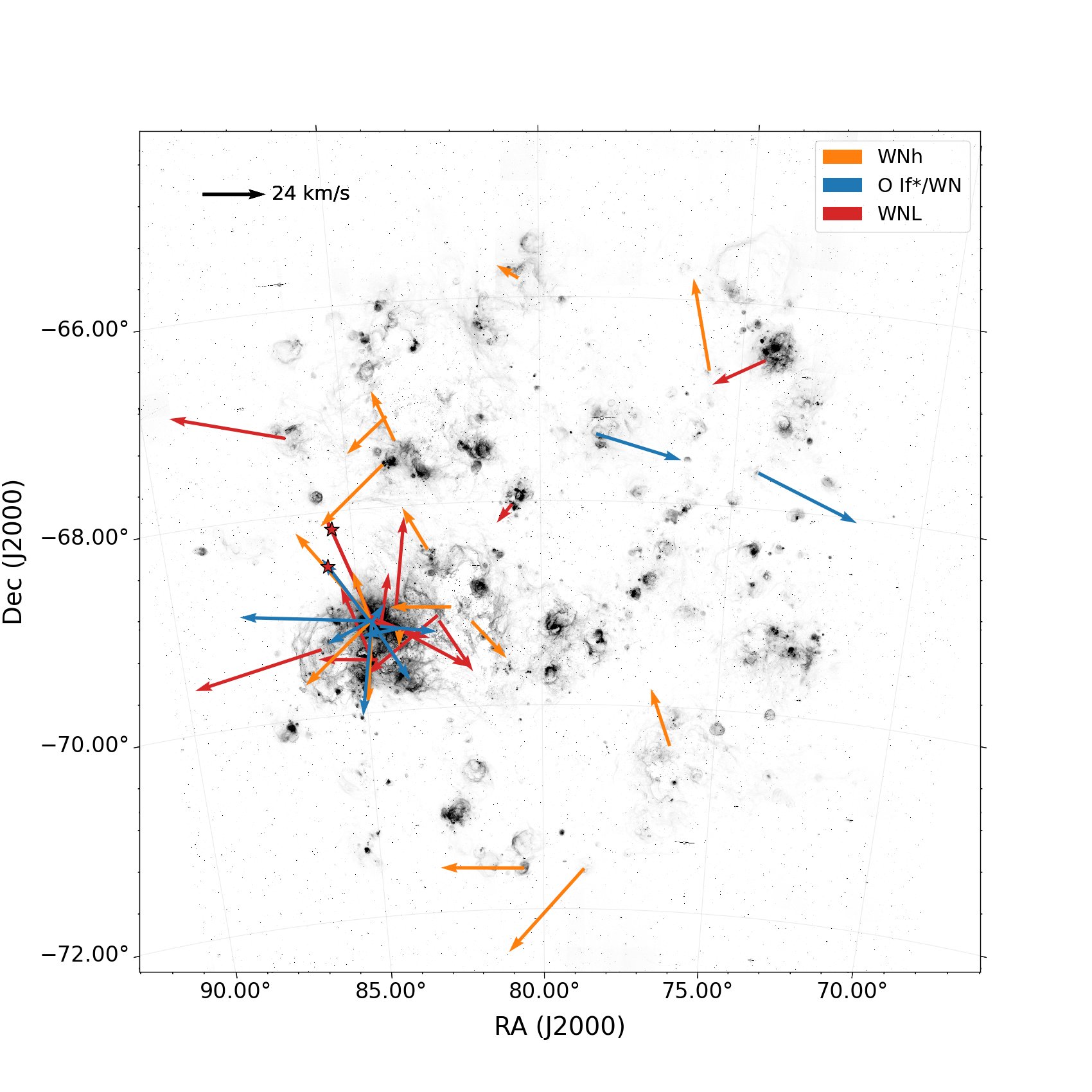}
    \caption{Proper motion vectors in the entire LMC for WNh, O If*/WN, and WNL stars. Panel ($a$) shows all stars with these subtypes, and panel ($b$) shows only the VMS stars, but additionally includes binary WNE stars, a class which also has high luminosities.  Plotting conventions are as in Figure~\ref{fig:vectors_30Dor}, including for the two WNL binary systems. }
    \label{fig:vectors_WN}
\end{figure*}

\setcounter{figure}{2}
\begin{figure*}
    \centering
    \includegraphics[height=7in]{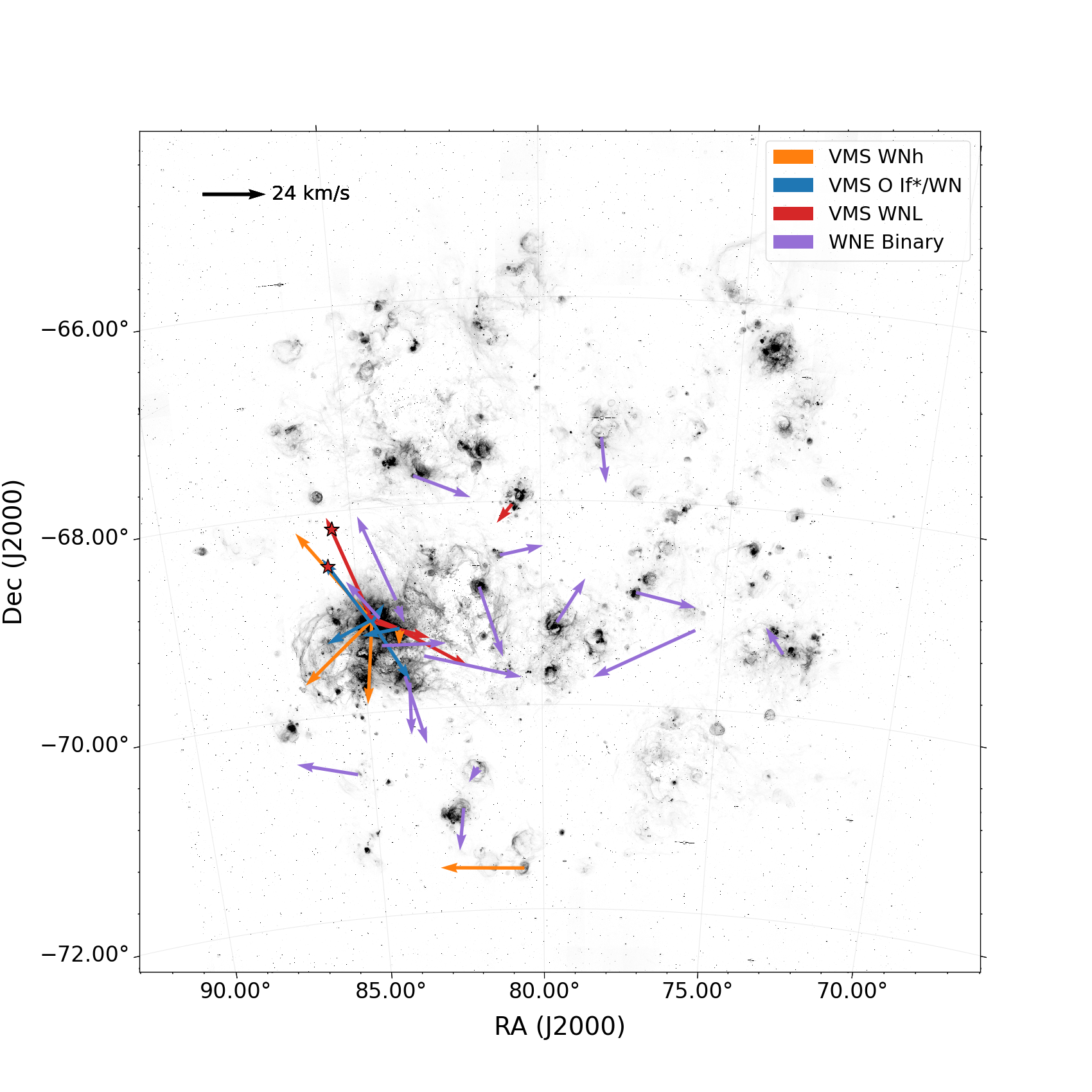}
    \caption{Figure~\ref{fig:vectors_WN}b}
\end{figure*}

\begin{deluxetable*}{lcccc}
\tablecaption{\hii\ region morphologies for various subpopulations}
\label{tab:HII}
\tablehead{
\colhead{\parbox[t]{2.75cm}{\centering Sub-population}} & 
\colhead{\parbox[t]{1.5cm}{\centering Total}} &
\colhead{\parbox[t]{1.5cm}{\centering Class 1}} & 
\colhead{\parbox[t]{1.5cm}{\centering Class 2}} &
\colhead{\parbox[t]{1.5cm}{\centering Class 3}} }
\decimalcolnumbers
\startdata
Total &  156 &  14 (9\%)\tablenotemark{a}&  72 (46\%)& 70 (45\%)\\
Single & 118 & 13 (11\%)\tablenotemark{a} & 47 (40\%) & 58 (49\%) \\
Binary & 38 & 1 (3\%) & 25 (66\%) & 12 (32\%) \\
Classical & 23 & 2 (9\%) & 4 (17\%) & 17 (74\%) \\
VMS  & 27 & 5 (19\%) & 22 (81\%) & 0 (0\%) \\
WN & 94 & 9 (10\%) & 39 (41\%) & 46 (49\%)  \\
\hline
WNh &  22 & 2 (9\%) & 10 (45\%) & 10 (45\%)  \\
WNh (Classical) & 11 & 0 (0\%) & 1 (9\%) & 10 (91\%)  \\
WNh (VMS) & 10 & 1 (10\%) & 9 (90\%) & 0 (0\%)  \\
\hline
O If*/WN &  13 & 2 (15\%) & 10 (77\%) & 1 (8\%)  \\
O If*/WN (Classical) & 3 & 1 (33\%) & 1 (33\%) & 1 (33\%)  \\
O If*/WN (VMS) & 8 & 1 (13\%) & 7 (88\%) & 0 (0\%)  \\
\hline
WNL & 22 & 4 (18\%) & 8 (36\%) & 10 (45\%)  \\
WNL (Classical) & 9 & 1 (11\%) & 2 (22\%) & 6 (66\%)  \\
WNL (VMS) & 9 & 3 (33\%) & 6 (67\%) & 0 (0\%)  \\
\hline
WNE & 60 & 3 (5\%) & 24 (40\%) & 33 (55\%)  \\
WNE Single & 39 & 2 (5\%) & 13 (33\%) & 24 (62\%)  \\
WNE Binary & 21 & 1 (5\%) & 11 (52\%) & 9 (43\%)  \\
\hline
WC &  23 & 3 (13\%)\tablenotemark{a} & 14 (61\%) & 6 (26\%) \\
WC Single & 13 & 3 (23\%)\tablenotemark{a} & 7 (54\%) & 3 (23\%) \\
WC Binary & 10 & 0 (0\%) & 7 (70\%) & 3 (30\%) \\
\hline
WN3/O3 & 10 & 0 (0\%)  & 3 (30\%) & 7 (70\%)  \\
WO & 3 & 0 (0\%) & 0 (0\%) & 3 (100\%) \\
\enddata
\tablenotetext{}{{\sc Note --} Data from \citet{Hung2021}. 
Stars classified as 1/2a in that work are assigned to Class 1 here.}
\tablenotetext{a}{Includes two stars classified as 1/2a in \citet{Hung2021}}
\end{deluxetable*}

Following \citet[][]{Hainich2014}, we define early-type WN stars (WNE) as those with WN classifications of WN2 -- WN5, and late-type WN stars (WNL) as types WN6 -- WN11. WN stars with composite O supergiant and WN spectra are known as O If*/WN, or 'slash stars,' while those that show some H in their spectra are classified as WNh stars. WNh and O If*/WN stars are excluded from both groups and analyzed separately, while Ofpe/WN9 stars are included within our WNL sample \citep{Crowther1995}.  Vector plots of the Gaia sample WNh, O If*/WN and WNL stars are shown in Figure~\ref{fig:vectors_WN}.

\subsection{VMS WNh, O If*/WN, and WNL Stars}\label{sec:VMS}

Many of the WNh, O If*/WN, and WNL stars are VMS with initial masses $\geq$ 100 $M_\odot$ \citep[e.g.,][]{Crowther2010}. As H-burning stars, VMS are likely to appear as WNh and O If*/WN stars, since these subtypes are characterized by the presence of atmospheric H, and many ordinary WNL stars may also originate from VMS progenitors \citep{Langer2012}. The extreme VMS luminosities can generate powerful stellar winds that lead to WN spectral features and allow the stars to self-strip much, but not necessarily all, of their outer, H-rich envelopes \citep[e.g.,][]{deKoter1997}. Thus, VMS stars are presumed to be able to reach WN phases as single stars, although they can also be formed as binary mass gainers and/or mergers. The VMS stars have been reported to differentiate from their lower-mass, ``classical" counterparts of the same subtypes at luminosities $\log L/L_\odot \gtrsim 5.9$ \citep{Pauli2022, Hainich2014, Crowther2011}. We therefore adopt this as the nominal threshold luminosity for VMS stars in what follows, and we find that the WNh, O If* and WNL with high luminosities seem to show distinct kinematics from the lower-luminosity, classical objects.

WNh stars are characterized by the presence of H in their spectra, as discussed by \citet{Smith2008}. The existence of WNh VMS stars have been well documented.  For example,  \citet{Crowther2010} show that the WN5h stars in the core of R136 have initial mass estimates $> 150\ M_\odot$, making them among the most massive stars known. Their stellar lifetimes are therefore among the shortest, with ages estimated to be 1.5 Myr \citep{Crowther2016}; stars at the top of the IMF tend to converge toward these short values as they contend with the Eddington limit \citep[e.g.,][]{Zapartas2017a}. 

The O If*/WN ``slash stars" show composite early O and mid- to late WN spectra. Most are also VMS and they are thought to be transitional between the optically thin winds of the most massive main sequence O stars and the optically thick winds of WN stars \citep{Vink2011}. As such, they may also be mostly single stars that self-strip their outer H-rich envelopes \citep[e.g.,][]{Crowther2011} or have other origins similar to those of VMS WNh stars. 

We note that ordinary WNL stars often show atmospheric H, unlike WNE stars, which tend to be H-poor.  This has caused some inconsistency in the literature regarding the classification of individual stars as WNh or not, and which stars are included in samples of WNh and O If*/WN versus ordinary WNL. For this analysis, we adopt the classifications in the \citet{Neugent2018} catalog. As noted above, our WNL sample excludes the objects in our WNh and O If*/WN samples.  These are also weighted toward earlier types; in particular, WN5h objects dominate the former.

VMS stars have the lowest frequency in the stellar initial mass function and would therefore preferentially form in the youngest, most massive clusters.  Their exceedingly short lives also imply that they have less time to travel from their birth clusters compared to other WR populations, and therefore they should still be found in close proximity to these clusters.  The spatial analysis by \citet{Smith2018} shows that WNh stars are among the least dispersed stars in the LMC, compared to other massive star populations. They show that about half of the WNh stars have the same spatial distribution as the earliest O stars, while the remainder are more dispersed.  This is consistent with the data in Table~\ref{tab:subpop} and Figure~\ref{fig:vectors_WN}b showing that all but 2 of our known WNh, O If*/WN, and WNL VMS stars are found in 30 Dor; we also note that there are 2 more stars with these subtypes of unknown luminosities in this region. Similarly, from the data of \citet{Hung2021}, we see that 100\% (with a lower limit of 73\%) of these VMS subtypes are found in Class 1 or 2 \hii\ regions (Table~\ref{tab:HII}), which are less evolved and less dynamically dispersed.  In contrast, comparing Figures~\ref{fig:vectors_WN}a and b, we see that the corresponding objects with classical luminosities show the opposite pattern and strongly favor locations outside 30 Dor. These data can be explained by the short VMS lifetimes which preclude such stars traveling far from their super star cluster birthplace, while classical WR stars form in a variety of lower-mass star-forming regions and have longer lifetimes, allowing them to travel farther from their parent clusters.

\begin{figure*}[ht!]   
    \includegraphics[width=1.0\linewidth]{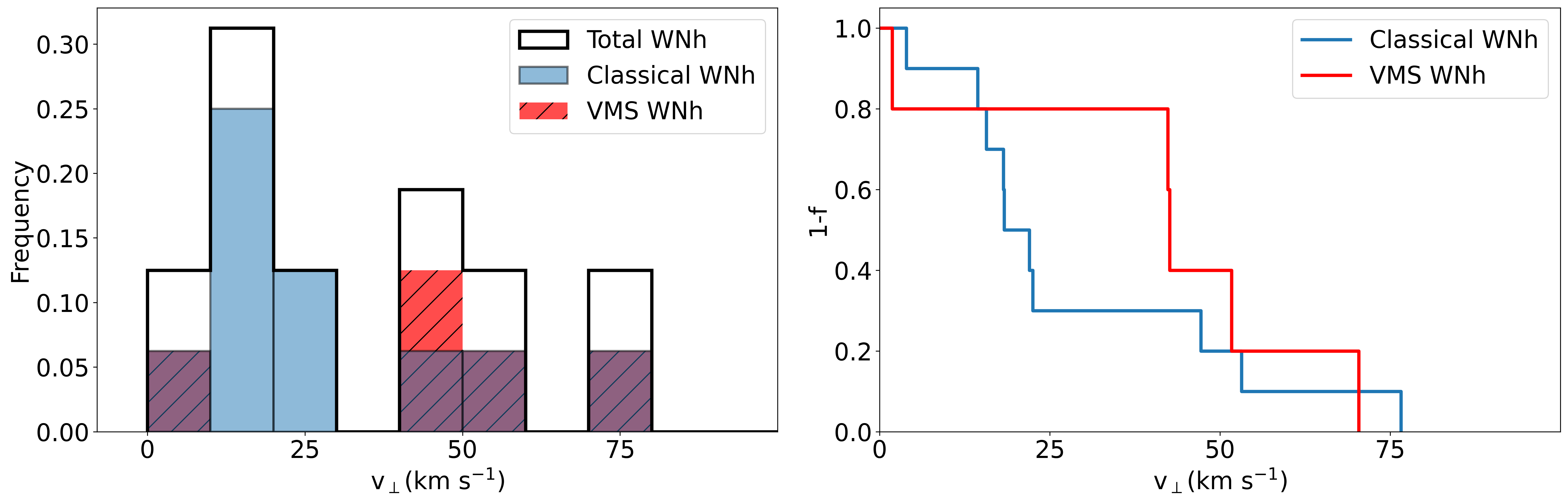}
    \includegraphics[width=1.0\linewidth]{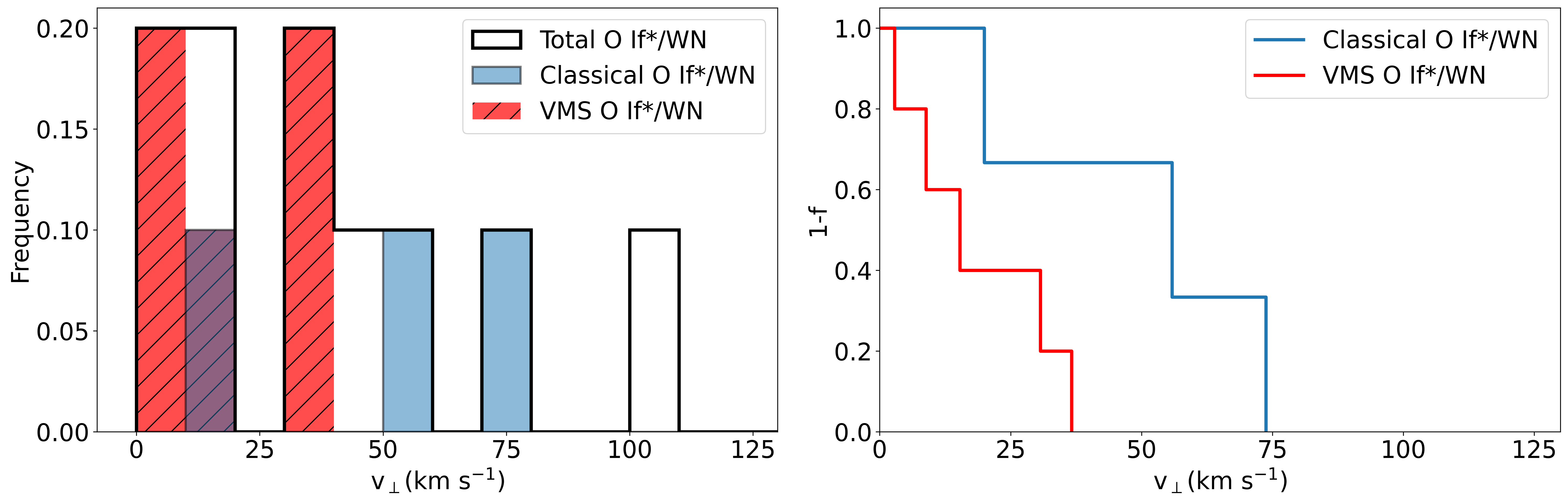}
    \includegraphics[width=1.0\linewidth]{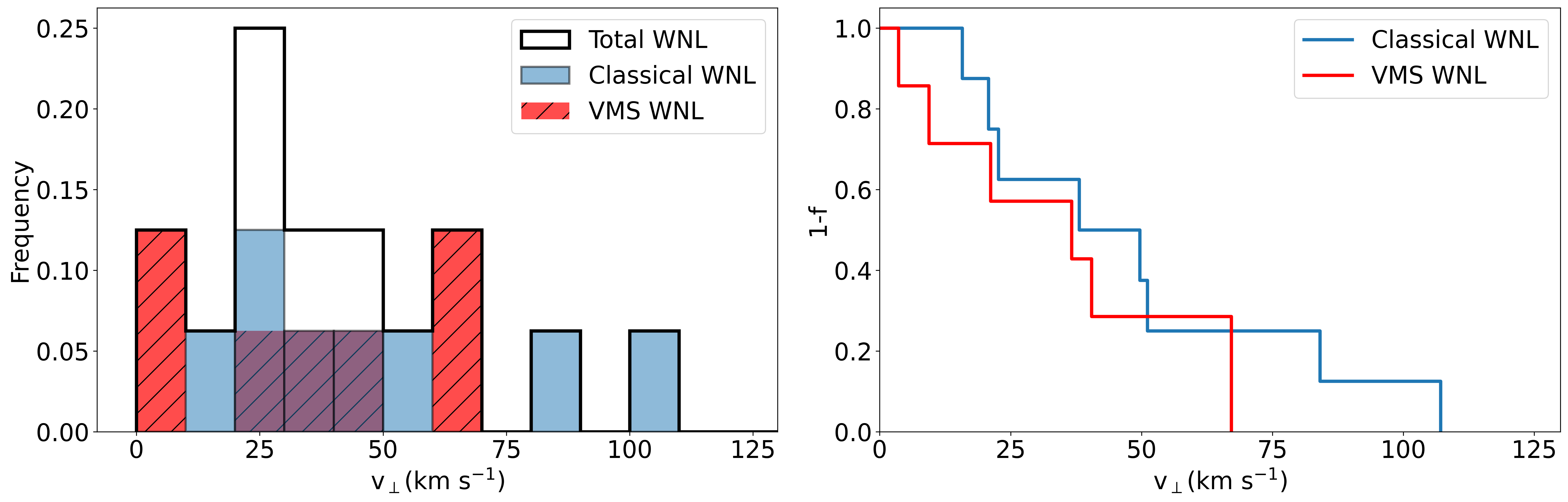}
    \caption{Left column:  Normalized differential transverse velocity distributions for WN stars with $L/L_\odot\leq 5.9$ (blue) and $L/L_\odot > 5.9$ (red). The black outlines show the distributions for all the subtype stars, including those for which no $L$ estimate is available.  WNh, O If*/WN, and WNL stars are shown in the top, middle, and bottom rows, respectively.  Right panel: Corresponding velocity survival functions $1 - f$, where $f$ is the cumulative distribution function, for the same samples.  The figures omit stars with no $v_\perp$ data: 6, 3, and 7 stars in the WNh, O If*/WN, and WNL samples, respectively. 
     \label{fig:veldist_WNh}}
\end{figure*}

\begin{figure*}
    \includegraphics[width=1.00\linewidth]{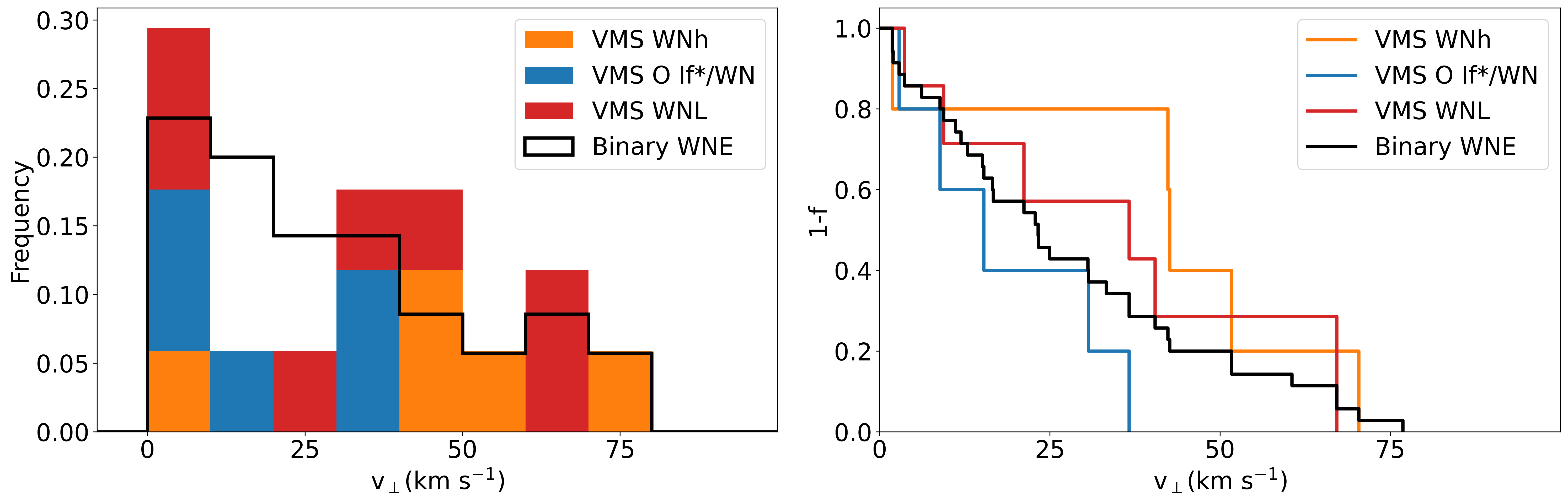}
    \caption{Left panel:  Normalized velocity distributions for VMS subtypes, colored as shown;  the black outline shows the overlapping distribution for binary WNE stars. The colored histograms are stacked and not overlapping.  
    Right panel:  Velocity survival functions for the same four groups. \\}
    \label{fig:veldist_VMSsub}
\end{figure*}

Figure~\ref{fig:veldist_WNh} shows the velocity distributions for VMS and classical stars in each of these subtypes.  We also show the corresponding $v_\perp$ survival functions $1-f$, where $f(v_\perp)$ is the cumulative velocity distribution function. In Figure~\ref{fig:veldist_VMSsub}, we show the combined distribution, which shows a significant component of slow objects with $v_\perp < 10$ \kms\ that has contributions from all three VMS subtypes, as well as a component of fast runaways.  This bimodality suggests that the slow objects with $v_\perp < 10$ \kms\ are dominated by unejected objects, while the ejected ones are most likely dominated by dynamical ejections.  BSS ejections, while possible, are unlikely since the short lifetimes of VMS \citep[$\lesssim2.5$ Myr; e.g.,][]{Martins2025} leave little time to observe them as post-SN binaries, and few are seen to be pre-SN binaries. The large number of runaway objects with $v_\perp > 24$ \kms\ (Figures~\ref{fig:veldist_WNh} and \ref{fig:veldist_VMSsub}) further support the dynamical mechanism, since SN ejections are generally not expected to accelerate surviving companions to such high velocities \citep[e.g.,][]{Renzo2019a, Brandt1995}; and moreover, the expected black hole kicks are weak for VMS \citep[e.g.,][]{VignaGomez2024}. It may be that VMS quickly sink to the cluster center due to mass segregation, causing them to be ejected early and faster while the cluster core is denser.

We note that the slow population is further enhanced by the objects in R136 that do not have velocity measurements, and are dropped from the sample due to crowding; such objects are likely unejected. This effect is most pronounced for the VMS O If*/WN stars, which have a median velocity of 8.9 \kms; this value is similar to that of Class 2 LBVs (median velocity 10 \kms), which \citet{Deman2024} suggest are similarly dominated by unejected objects. Table~\ref{tab:subpop} shows that O If*/WN stars have the highest frequency of stars in 30 Dor. Among all the LMC O If*/WN stars, a majority (11/13 or 85 $\pm$ 35\%) are found in the 30 Dor region, and of these, 7 out of 11, or 64 $\pm$ 31\% are found in the R136 super star cluster. For the VMS WNh stars, there are 4 with no Gaia measurements in R136 and 1 slow star, while 3 stars are fast runaways with velocities $> 40$ \kms, and 1 remaining star outside R136 has no Gaia data.  This is again consistent with a bimodal velocity distribution, although we caution that these are small-number statistics. The WNL stars show a broader range of velocities, perhaps with kinematics intermediate between the other two subtypes, supporting the premise that they are closely related and consistent with some of the stars with these subtypes being interchangeable with WNL types. 

The total number of VMS in R136 omitted from our Gaia sample correspond to 4 VMS WNh stars and all 3 of the dropped VMS O If*/WN stars. These 7 stars may belong in the lowest velocity bins in the Figures~\ref{fig:veldist_WNh} and \ref{fig:veldist_VMSsub}, suggesting that up to half of the VMS may be unejected.  This is consistent with the finding by \citet{Stoop2024} for the R136 massive-star population in general.
The data would therefore imply that {\it the ejection timescale is similar to the ages of these VMS types,} and it supports the predicted ejection timescale from dense star clusters \citep[e.g.,][]{Oh2015, Oh2016} and predicted ages for the R136 VMS \citep{Crowther2016} which are both on the order of 1.5 Myr.

\begin{figure*}
    \includegraphics[width=1.00\linewidth]{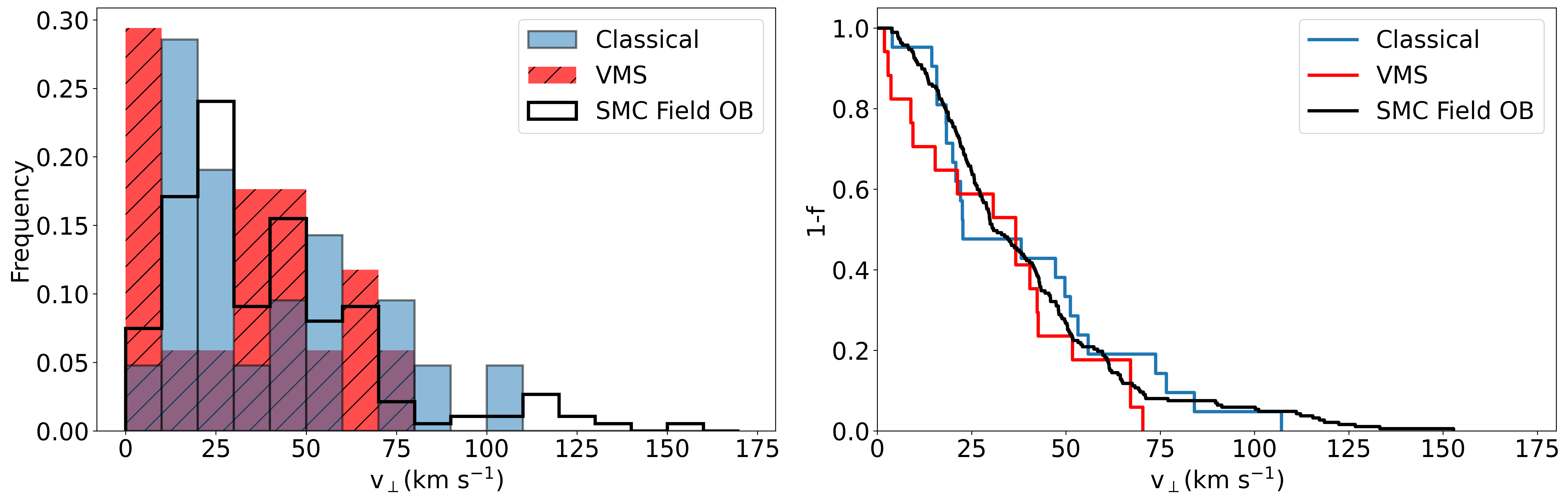}
    \caption{Left panel:  Normalized velocity distributions for the combined WNh, O If*/WN, and WNL stars with 
    lower, classical luminosities ($L/L_\odot\leq 5.9$, blue) and higher, VMS luminosities ($L/L_\odot > 5.9$, red).
    Right panel:  Velocity survival functions for the same two groups.
    Additionally, both panels show the distribution for the SMC field OB stars from \citet{Phillips2024} overplotted with black lines.
    % ; the additional green line shows the high-$L$ sample combined with binary WNE stars. 
    \\}
    \label{fig:veldist_VMS}
\end{figure*}

The SMC field OB stars have been suggested to represent a population of massive stars whose kinematic footprint is also believed to be dominated by DES ejections \citep{Phillips2024}. Figure \ref{fig:veldist_VMS} compares these two velocity distributions.  Both the SMC field OB stars and the LMC VMS WR stars show distributions with multiple-peak structure, including similarities in the high-velocity regime. While both populations may have contributions from multiple ejection mechanisms, we argued above that dynamical ejections play a major role in accelerating VMS WR stars, thus the similarities with the field OB star velocities support the scenario that they, too, include a dominant population similarly ejected by the DES as suggested earlier \citep[e.g.,][]{Phillips2024, DorigoJones2020}. 

Figure~\ref{fig:veldist_VMS} does show a significant walkaway component with $v_\perp \lesssim 24$ \kms\ among the field OB stars that is not observed for the VMS WR stars.  We note that differences between these two populations are expected, since VMS WR stars and OB stars originate in different areas of the upper IMF, and VMS stars are on average much more massive than typical OB stars.  Moreover, the parent clusters of the LMC VMS WR stars and SMC OB stars also have different average properties, including the unique role of R136 in ejecting the VMS WR stars (Figure~\ref{fig:vectors_WN}b). It is unclear whether the OB walkaway stars are due to such differences, or to a different mechanism altogether.

\begin{figure}[h!]
    \centering
    \includegraphics[width=1.0\linewidth]{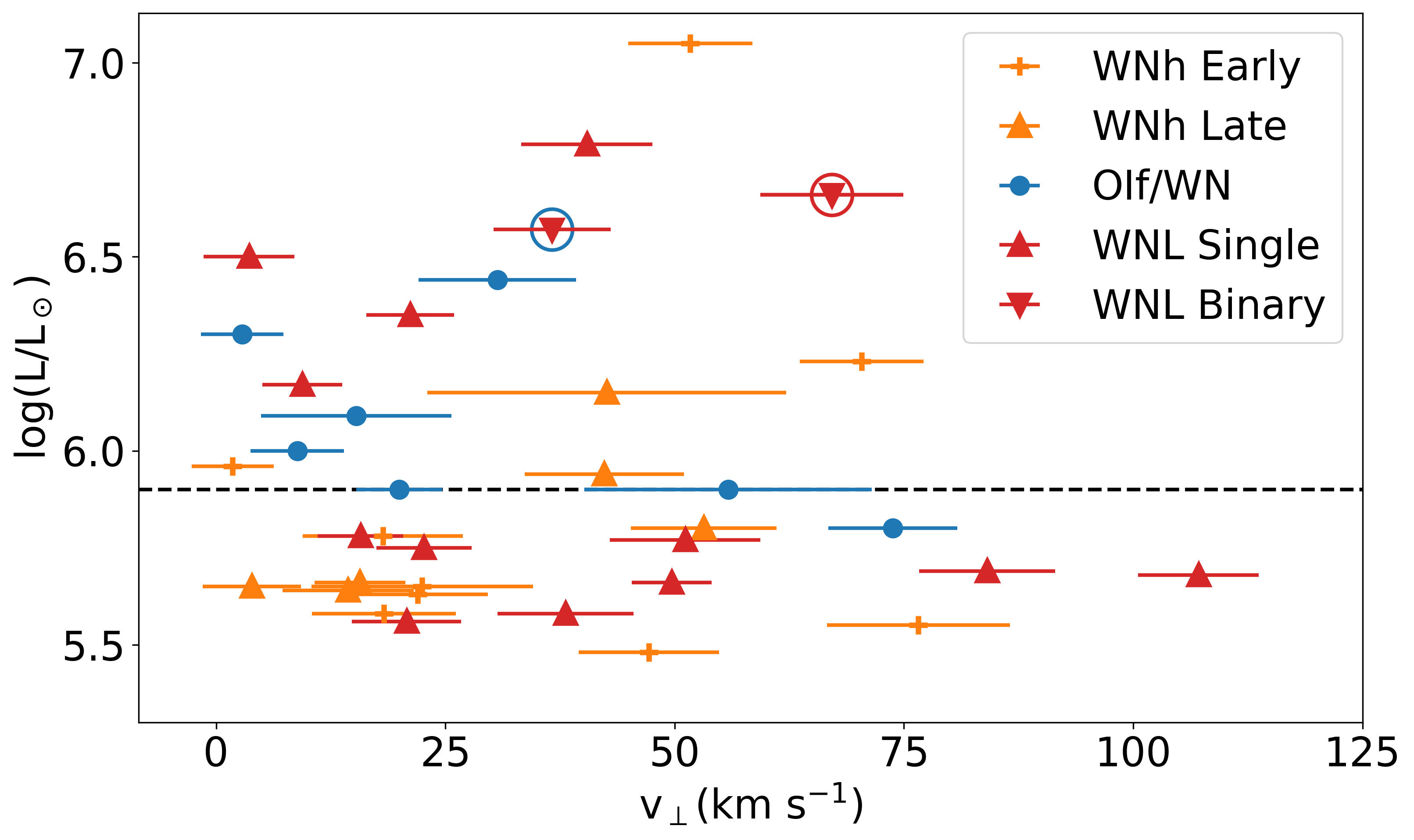}
    \caption{Luminosity vs $v_\perp$ for WNh, O If*/WN, and WNL single and binary stars, as shown, for objects with available luminosities from \citet{Hainich2014}.  The dashed line indicates the threshold luminosity $\log L/L_\odot = 5.9$ adopted for VMS status.  Objects with circles are binaries, with the circle color corresponding to the companion subtype.}
    \label{fig:ScatterVMS}
\end{figure}

Almost all of the WNh, O If*/WN, and WNL VMS are apparently single stars.  Their binary fractions are, respectively, $10\pm10$\% (1/10),  $13\pm13$\% (1/8), and $\sim22\pm17$\% (2/9, or 3/9 when counting both stars in the BAT99-118 WN5/6 + WN6/7 binary), with the caveat that the remaining VMS WNL stars may have undetected companions \citep[][]{Hainich2014}. The two pre-SN binaries have among the highest velocities and luminosities among the WNL stars (Figure~\ref{fig:ScatterVMS}), and these two objects must be dynamically ejected. 
The VMS could originate as single stars; \citet{Schneider2018} show that the 30 Dor OB population is compatible with an IMF of single stars extending up to at least 200 $\msun$. Alternatively, they could be binary mass gainers or mergers \citep[e.g.,][]{Crowther2011, Shenar2019a}, which would spin up their rotation velocities; \citet{Roy2020} show that massive stars rotating at $\gtrsim$ 30\% of their breakup velocity will evolve WNL-like surface abundances, which could support such a scenario. Although it is difficult to determine WR rotation velocities due to the lack of atmospheric features \citep[e.g.,][]{Renzo2019b}, many WNh stars have been shown to be rapid rotators \citep[e.g.,][]{Crowther2010, Martins2009}. It may be likely that there are multiple routes to generating VMS, and \citet{Pauli2022} find substantial contributions from both binary and single-star formation in this regime, based on their stellar population synthesis models.

\subsection{Classical WNh, O If*/WN, and WNL Stars}

The lower-luminosity, classical WNh, O If*/WN, and WNL stars with $\log{L/L_\odot} < 5.9$ are thought to be lower-mass stars having ages $>2.5$ Myr, which may be H or He-burning \citep[e.g.,][]{Sen2023, Pauli2022}. In contrast to the VMS stars, they are found throughout the LMC and only 4 of the 23 known classical stars are in 30 Dor (Table~\ref{tab:subpop}), indicating that they largely originate from lower-mass clusters throughout the galaxy. \citet{Hainich2014} find that classical WNL stars are dominated by single stars, while the high-$L$ counterparts are not, further supporting their likely different nature and origin. However, \citet{Pauli2022} predict that it is the high-$L$ objects that should originate from single-star evolution and the lower-$L$ from binaries, which is opposite to this finding.  While binaries can be hard to detect, the mass-gainer companions in the \citet{Pauli2022} models are predicted to be more luminous than the WR stars, which seems difficult to explain in this scenario. 

Table~\ref{tab:HII} shows that the \hii\ regions associated with classical WNh, O If*/WN, and WNL stars are strongly dominated by Class 3, which are the most evolved and dispersed nebular type in the classification scheme of \citet{Hung2021}.  This is consistent with their observed velocities (Figures~\ref{fig:veldist_VMS} and ~\ref{fig:ScatterVMS}), which are also dominated by values implying ejections from clusters.  The data demonstrate that these classical WR stars are more spatially dispersed than the VMS stars, consistent with their older expected ages.

Comparing the total, combined peculiar velocity distributions for the classical and VMS WNh, O If*/WN, and WNL stars, both distributions show significant structure with multiple peaks (Figure~\ref{fig:veldist_VMS}). The classical stars lack the slow, unejected component, and are instead strongly dominated by ejected stars; among all three subtypes, there is only 1 classical WNh star in 30 Dor lacking a Gaia velocity measurement. The velocity peaks appear to complement those of their VMS counterparts, and appear to be shifted to higher values in Figure~\ref{fig:veldist_VMS}.  The qualitative similarity suggests that the classical objects are similarly dominated by dynamical accelerations.  The offsets in velocity may be linked to their lower masses and/or lower-mass birth clusters. Ejection velocities scale primarily with the binding energy, i.e., mass of the encountered binary system, whether in a ``flyby" or ``exchange" interaction \citep[e.g.,][]{Banerjee2012}, and binaries in small clusters are systematically less massive when there is stochasticity in the upper IMF \citep[e.g.,][]{Oey2005}.  On the other hand, \citet{Farias2019} and \citet{Farias2023} predict that lower-mass clusters eject higher-velocity O stars when the star-formation efficiency is low, allowing stars to remain longer in high-density groups, and thereby facilitating fast ejections.  Thus, the extent to which runaway velocities vary with cluster mass is unclear, as is the relationship between the VMS and classical velocity distributions.

The classical population also includes a kinematic component with walkaway velocities in the 10 -- 30 \kms\ range.  As discussed above, this component is seen for the SMC field OB stars, but not the VMS. This complex velocity distribution may again point to multiple acceleration mechanisms.

\citet{Sen2023} suggest a ``reverse Algol" scenario for generating this population of lower-$L$, H-rich WN stars, in which
the primary remains more luminous and massive than the secondary in a semi-detached system. Case A mass transfer takes place on the main sequence, and the more massive donor star would develop a WR wind with optical depth similar to that of observed WNh, O If*/WN, and WNL stars while still core H-burning, or shortly thereafter. These donors, being more massive and luminous, may hide their companions and appear to be single stars. Due to the relation between wind mass-loss rate and luminosity, these massive donor stars could develop strong winds at masses lower than would be expected from VMS MS evolution driven by wind self-stripping. As binaries, ejected objects formed in this way would be dominated by the dynamical mechanism.

\section{WNE Stars}

\begin{figure*}
    \centering
    \includegraphics[height=7in]{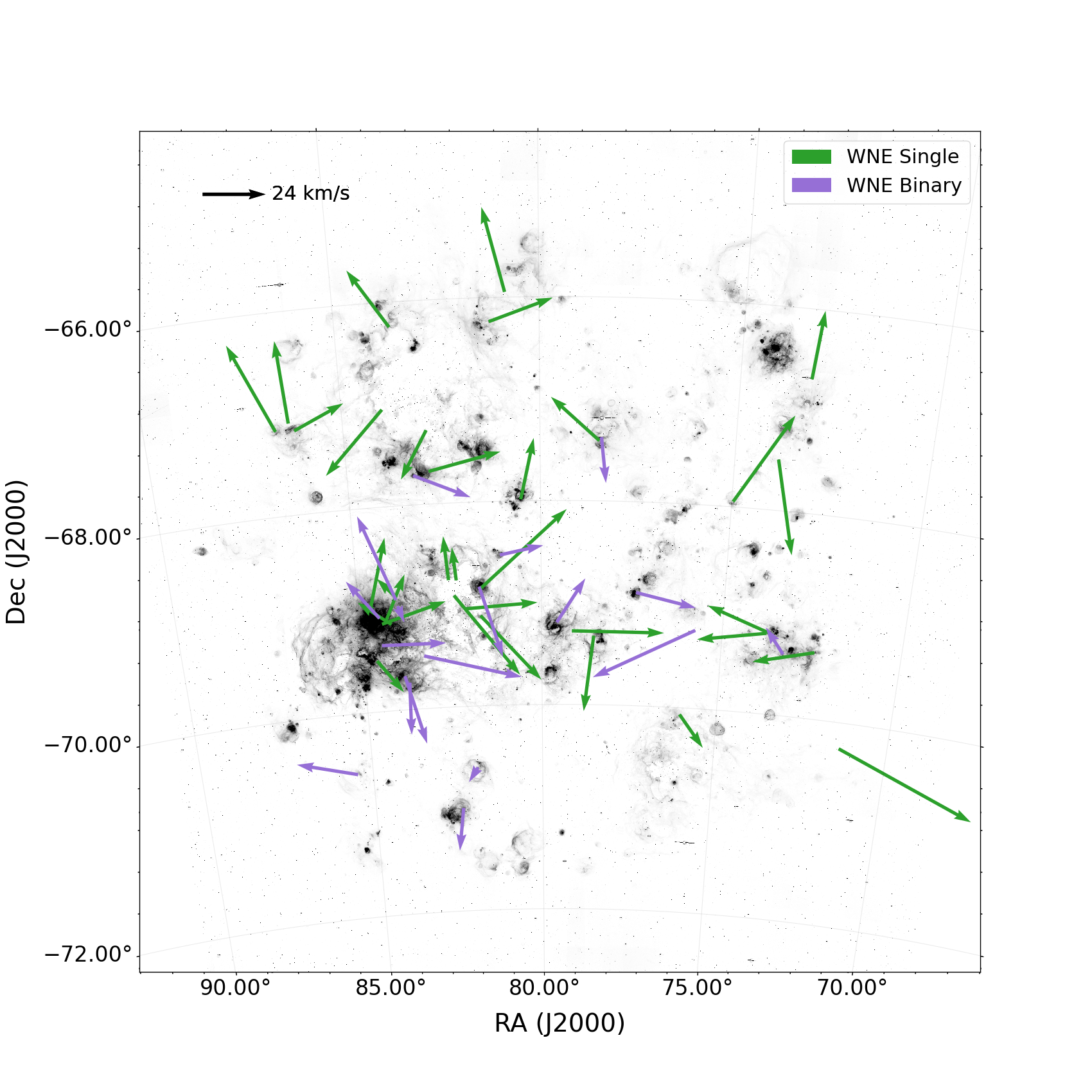}
    \caption{Same as Figure~\ref{fig:vectors_WN}, but for single and binary WNE stars.}
    \label{fig:vectors_WNE}
\end{figure*}

\begin{figure*}
    \centering
       \includegraphics[width=1.00\linewidth]{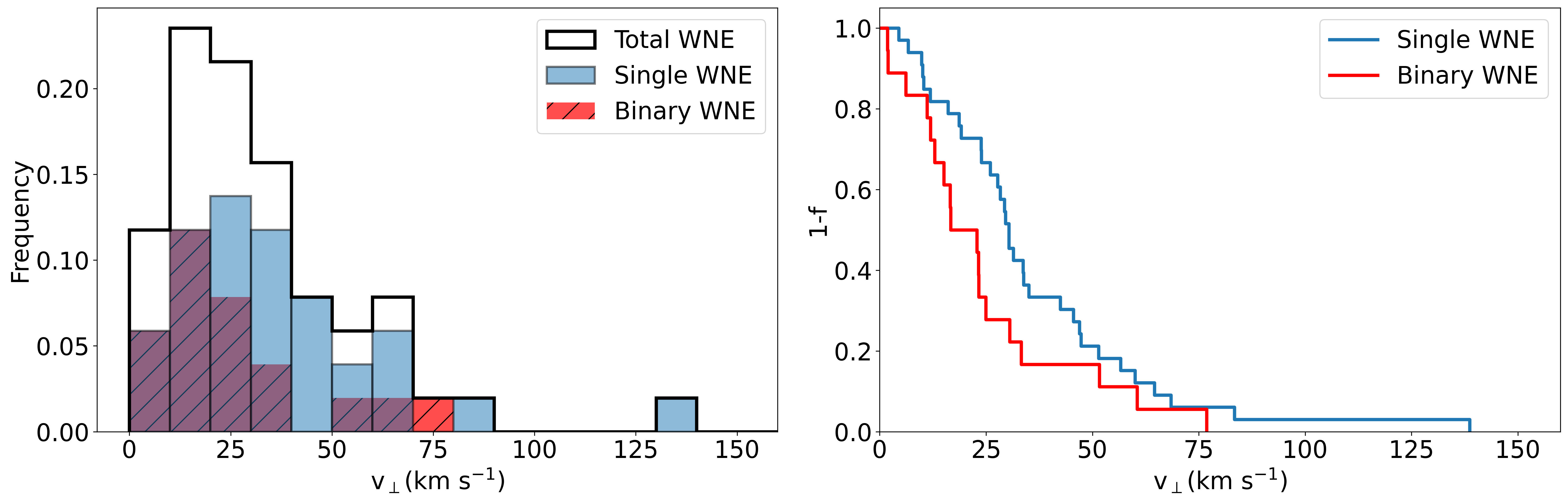}
    \caption{Same as Figure~\ref{fig:veldist_WNh}, but for WNE stars, showing that the single stars are dominated by runaways and the binaries are dominated by walkaways. \\}
    \label{fig:WNE_vel}
\end{figure*}

WNE stars represent a lower-mass WR population, and as such, they provide the largest contribution to both the total (60/156 or $38\pm6$\%) and Gaia (51/121 or $42\pm7$\%) samples. Our WNE samples exclude the 15 early WNh and O If*/WN stars, at least 5 of which have VMS luminosities, and also the WN3/O3 stars (Section~\ref{sec:wnodiscussion}). \citet{Hainich2014} show that WNE stars have a median mass of $\sim 25\  \msun$, with luminosities $ \log(L/L_{\odot}) \sim 5.3-5.8$, both of which are lower than their WNL counterparts. The nebular environments of the LMC WNE stars are dominated by the more evolved, Class 3 \hii\  regions (Table~\ref{tab:HII}), consistent with both longer lifetimes and longer travel times. As noted above, WNE stars are also more H-depleted ($X_H \lesssim 0.05$) and so they could originate as either post-LBV or post-RSG helium-burning objects, with progenitor initial masses of $\sim60-25\ M_{\odot}$ in the Milky Way \citep[e.g.,][]{Langer2012}.  In the LMC, \citet{Hainich2014} constrain their initial masses to $< 40\ M_{\odot}$. Although WNE stars are a large population, only 8 ($13\pm5$\%) of them are found in 30 Dor (Figure~\ref{fig:vectors_WNE}), implying that the currently observed population originates from other, lower-mass clusters with a range of ages.  This is consistent with their lower stellar masses and relatively large number. 

Binary processes can be expected to be important for the origin of WNE stars, especially since the LMC's lower metallically may mean wind stripping will be weaker \citep[$\sim Z^{0.5}$][]{Kudritzki1989}, although as noted above, the threshold for WR phenomena is also metallicity-dependent \citep{Shenar2020}. This is supported by the high observed binary frequency of WNE stars. Among sub-populations of WR subtypes in our total sample, they have the second-highest fraction, $35\pm9$\% (21/60) of non-compact, pre-SN binaries, second only to WC stars. Binary WNE stars have a median $v_\perp$ of 17 \kms, while single WNE stars have a median $v_\perp$ of 30 \kms. The velocity distributions for the single and binary WNE stars are shown in Figure \ref{fig:WNE_vel}.  A two-sided K-S test comparing the two groups yields a test statistic of $D=0.39$ and a $p$-value of 0.04, which shows that the difference in their velocity distributions is statistically significant. This may suggest that they have different origins, and that the single WNE stars are not the survivors of the binary WNE systems.  Such a scenario where binary and single WR stars appear to have different origins is apparently the case for WC stars (see Section~\ref{sec:wcdiscussion} below).  On the other hand, the shapes of the two velocity distributions in Figure~\ref{fig:WNE_vel} are qualitatively similar, other than a general offset to lower velocities for the binary stars.

\begin{figure}
    \centering
    \includegraphics[width=1.0\linewidth]{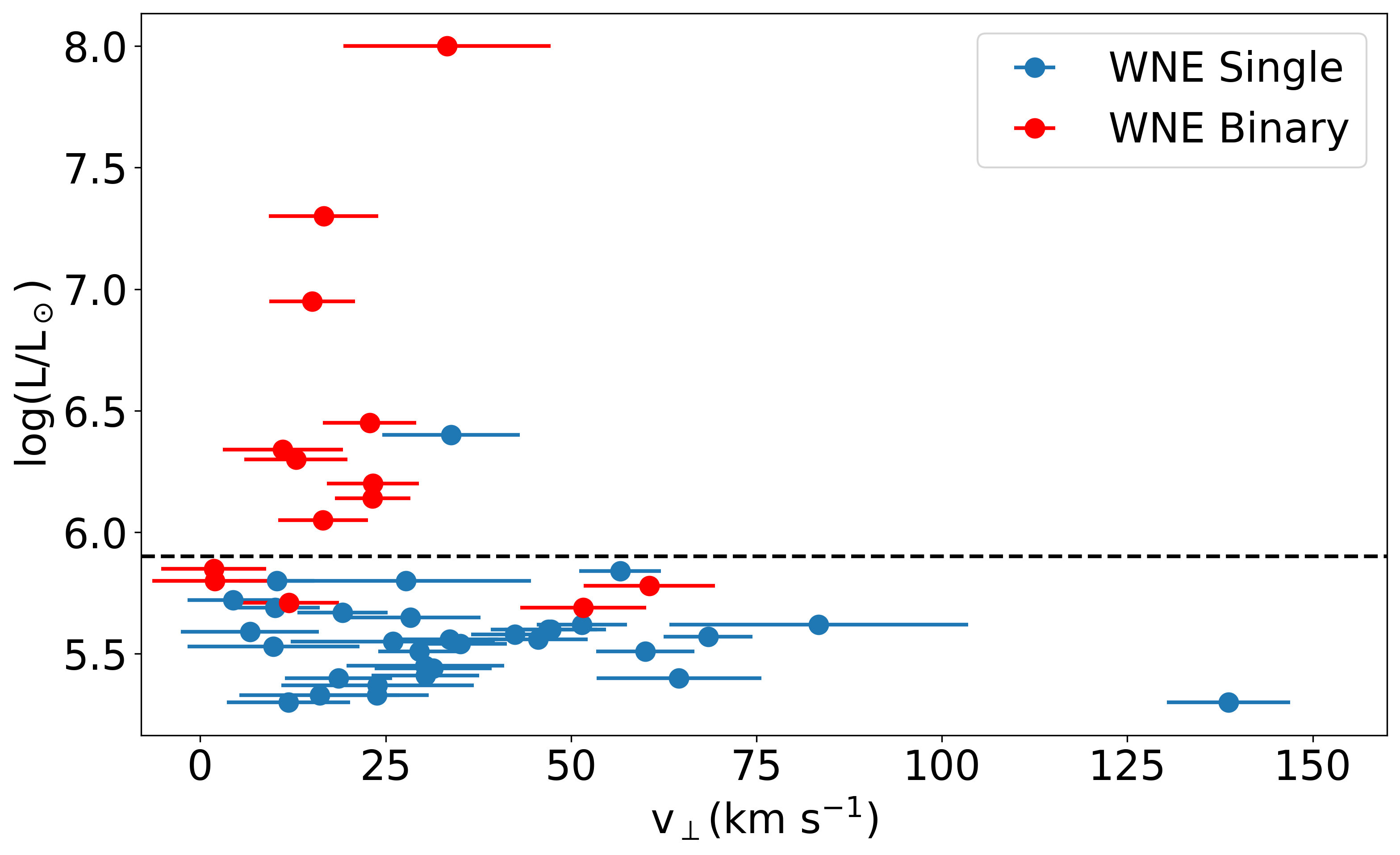}
    \caption{Same as Figure~\ref{fig:ScatterVMS} for single and binary WNE stars.} 
    \label{fig:ScatterWNE}
\end{figure}

Figure~\ref{fig:ScatterWNE} shows that the binary WNE stars are systematically more luminous than the single ones, and are mostly consistent with VMS luminosities. This trend is again opposite to that predicted by \citet{Pauli2022}, who find that the lower-$L$ WR stars should be binaries, including the H-free WN stars. However, none of the binary WNE stars are associated with R136, and Figure~\ref{fig:vectors_WNE} shows that they originate from clusters all over the LMC.  Whereas 0\% of VMS are found in Class 3 \hii\ regions, this applies to $43 \pm 17\%$ of the WNE binaries (Table~\ref{tab:HII}). Figure~\ref{fig:veldist_VMSsub} also shows that the shape of the velocity distribution for WNE binaries is apparently different from that of the VMS stars. A one-sided K-S test between the two groups yields $D=0.37$ and a $p$-value of 0.08 for the probability that the WNE binaries are faster than the VMS stars. The binary WNE kinematics are more similar to those of the single WNE stars (Figure~\ref{fig:WNE_vel}) and the OBe stars in Figure~\ref{fig:OBevsWNE} below.  It therefore seems likely that the binary WNE stars either have luminosities that are overestimated due to the approximate method used to derive them \citep{Hainich2014}, or their luminosities are enhanced by binary interactions, for example, due to inflated, fast-rotating envelopes from mass transfer \citep[e.g.,][]{Castro2018}, or as mass gainers that attained VMS masses.

\begin{figure}
    \centering
    \includegraphics[width=1.00\linewidth]{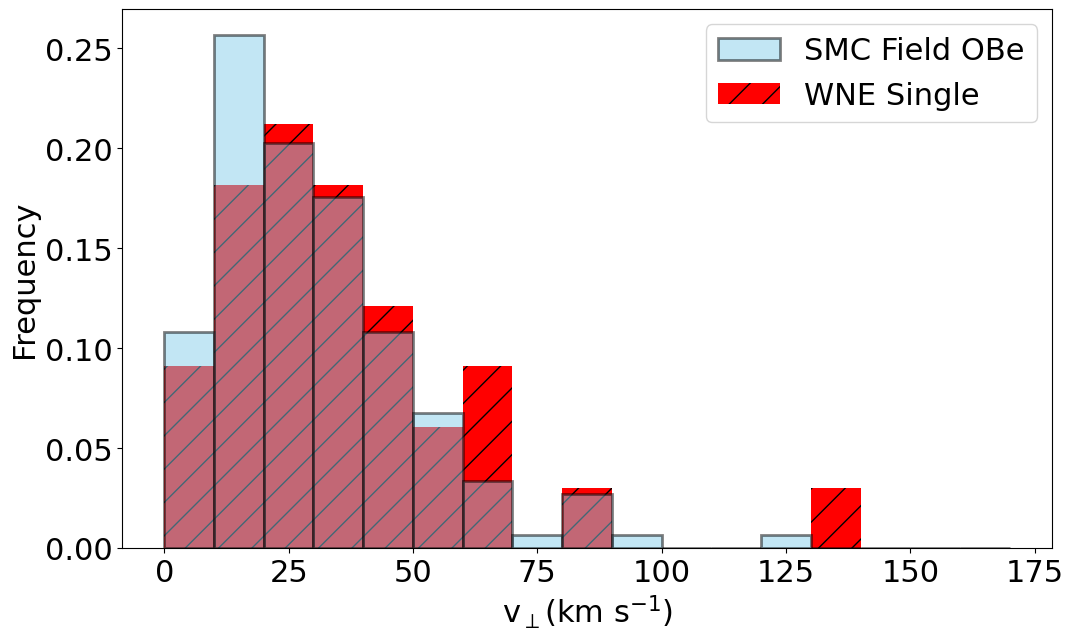}
    \caption{Normalized velocity distributions of LMC WNE stars vs SMC OBe stars}
    \label{fig:OBevsWNE}
\end{figure}

\subsection{Single WNE stars: Possible mergers?}\label{sec:merger}

The single-star WNE velocity distribution seems to differ in character from those of the WN subtypes we examined in Section~\ref{sec:WNLtypes}.  Figure~\ref{fig:OBevsWNE} shows a single dominant peak near the median $v_\perp$ of 30 \kms, whereas the other subgroups have distributions suggesting more complex structures (e.g., Figure~\ref{fig:veldist_VMS}).  Instead, the velocity distribution of single WNE stars is qualitatively similar to that of SMC field OBe stars from \citet{Phillips2024}, which are also shown in Figure \ref{fig:OBevsWNE}, although the OBe stars peak at a lower velocity (median 24 \kms).  OBe stars are believed to be mass gainers from binary systems that have been accelerated by SN kicks, rather than dynamical ejections. While the similarity with OBe stars may be a coincidence, the shape of the velocity distribution for single WNE stars suggests an acceleration mode that is fundamentally different from that of the VMS and other H-rich WN stars, in particular, binary-driven ejection rather than dynamical ejection. 

If single WNE stars are accelerated through the SN mechanism, then somehow the stripped primary in the binary must be the survivor. \citet{Pols1994} suggested a rare scenario in which the secondary star gains enough mass to explode first.  However, the large population of single WNE stars seems inconsistent with this possibility, and it may better apply to WC or WN3/O3 stars (see Sections~\ref{sec:wcdiscussion} and \ref{sec:wnodiscussion} below).  Instead, we speculate that single WNE stars may be the result of mergers in which the primary is stripped via a common envelope phase followed by merger phase.

Most massive stars are born with a close companion \citep{Sana2012,Offner2023}, and a significant fraction are expected to undergo a merger during their lifetime \citep{Sana2012, Renzo2019a}, by which we mean a failed common-envelope event where the envelope is not entirely ejected and no binary survives. Substantial areas of the parameter space in binary period and mass ratio are expected to generate mergers \citep[e.g.,][]{Pauli2022}, and in particular, systems with low initial mass ratios $q_i$.  {\it Thus, mergers present another, relatively unexplored mechanism for generating runaway stars.} WR merger descendants could also alleviate an overprediction of binary WR stars suggested by \citet{Pauli2022}. \citet{Li2024} explore the possibility mergers evolving into WR stars, suggesting that the  overluminous and/or rapidly rotating product is less bound and therefore easier to self-strip.  Here, we also suggest that binaries with low $q_i$ could provide another scenario for generating WR merger descendants by explosive stripping as follows.  

During a merger of two stars, some mass always gets ejected \citep{Ivanova2016,Ivanova2020}. This initial  ejecta, produced before the companions come into physical contact, carries away a significant amount of the initial orbital angular momentum and is highly asymmetric, unlike the subsequent ejecta phases in a common-envelope event that leads to binary survival. Hydrodynamical simulations indicate that the initially ejected fraction of the donor mass, $f_{\rm ej}$, is usually between 2\% and 15\% of the total donor mass $M_{\rm d}$ \citep{Glebbeek2013,Ivanova2016}.  The material is expelled with a velocity $v_{\rm ej}$ ranging from about the donor's escape velocity, $v_{\rm esc}$, up to $3v_{\rm esc}$. Because the ejecta remove a substantial amount of angular momentum asymmetrically, the merger remnant receives a recoil or ``kick.'' The upper limit on this recoil velocity, to order of magnitude, can be estimated from momentum conservation as

\[ v_{\rm rec} = v_{\rm ej} \frac{f_{\rm ej}M_{\rm d}} {(1-f_{\rm ej}) M_{\rm d} + M_{\rm c} }\ , \]
where $M_{\rm c}$ is the companion mass. For example, a 25 $M_\odot$ donor of 100 $R_\odot$ merging with a companion ten times less massive would experience a recoil of $\sim$30 km s$^{-1}$ if $f_{\rm ej}=0.05$ and $v_{\rm ej}=2v_{\rm esc}$, and up to $\sim$150 km s$^{-1}$ for $f_{\rm ej}=0.15$ and $v_{\rm ej}=3v_{\rm esc}$. Binary population synthesis modeling is needed to estimate the resulting velocity distribution, but these fiducial values are consistent with the observed WNE velocities.  

If part of the envelope survives, the subsequent evolution of this hydrogen-rich material becomes critical. When a relatively unevolved main-sequence companion merges with an evolved massive star that has a core and burning shells, an additional mass-transfer phase may occur. Specifically, as the plunge continues, the companion can approach the core closely enough to overflow its instantaneous Roche lobe and transfer mass directly onto the core and surrounding burning shells, whether the H-burning shell alone, or both the H- and He-burning shells \citep{Ivanova2002b,Ivanova2002a,Ivanova2003}. If the companion is of low mass ($\lesssim2M_\odot$), its low-entropy, hydrogen-rich material may penetrate into the He-burning shell, triggering explosive hydrogen burning and removing not only the envelope but also the He-burning shell. In outcomes where the stripping is sufficiently deep, the exposed He- and N-rich layers cause the remnant to present itself observationally as a hydrogen-poor, WNE star after the merger.

An intermediate-mass companion ($3-7\ M_\odot$) does not cause shell explosions but can dredge up He-rich core material and increase the envelope mass, often transforming the star into a blue supergiant \citep{Ivanova2002a,Menon2017}. Although less dramatic than the low-mass case, this pathway can still culminate in WNE formation once subsequent winds and rotationally enhanced mass loss peel away the remaining envelope \citep{Li2024}. He-rich material is dredged up, enriching the new envelope in helium and CNO elements. The He- and CNO-enriched envelope also rotates much more rapidly than before the merger.

This scenario for the origin of the observed LMC WNE stars would suggest that the SMC OBe stars could also be mergers, but with surviving H envelopes, perhaps originating from systems with higher $q_i$.  This might potentially explain their qualitatively similar velocity distributions (Figure~\ref{fig:OBevsWNE}).

These post-merger properties include higher effective temperature at fixed luminosity, enhanced CNO abundances, and rapid rotation. Higher effective temperature often increases the mass-loss rate \citep{Vink2000}, CNO enrichment tends to enhance line-driven winds \citep{Vink2005,Grafener2008,Krticka2009}, and rapid rotation further strengthens wind mass loss near Eddington conditions \citep{Langer1998,Maeder2000}, together accelerating envelope stripping. In some cases, the inflated, rapidly rotating, chemically enriched envelope may also approach the Eddington limit, triggering LBV-like instabilities and episodic eruptions that further hasten envelope removal \citep{Justham2014,Vink2018, Cheng2024}.

\section{WC Stars}
\label{sec:wcdiscussion}

\begin{figure*}[h!]
    \includegraphics[width=1.0\linewidth]{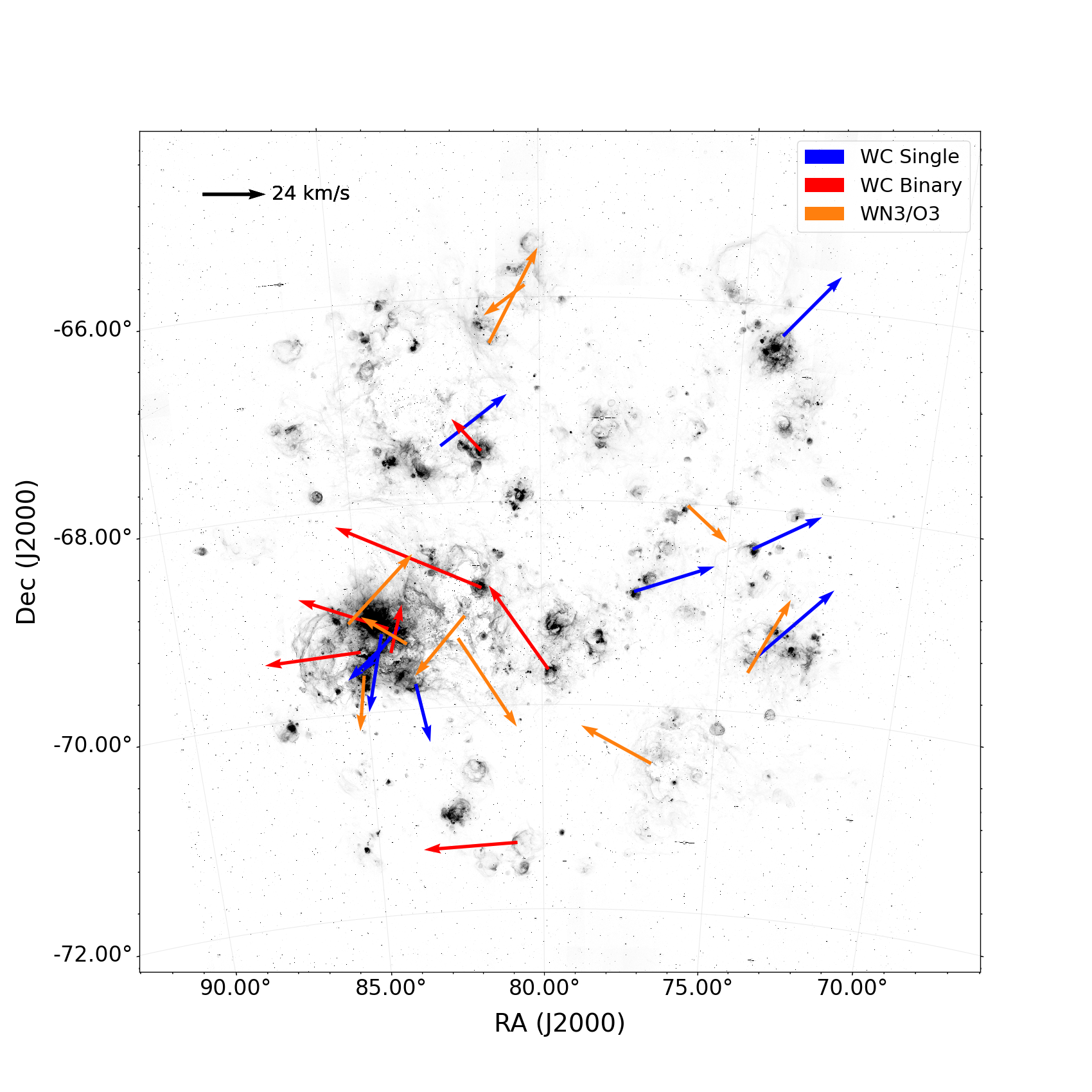}
    \caption{Vector plot showing $v_\perp$ for single (blue) and binary (red) WC stars, and WN3/O3 stars (orange) overlaid on an H$\alpha$ image from \citet{Smith2005}. Vectors are scaled as $v_\perp^{1/2}$.}
    \label{fig:vector_WC}
\end{figure*} 

WC stars are thought to represent a more stripped and evolved WR phase relative to the WN stars \citep[e.g.,][]{Langer2012}. The WC stars are characterized by emission lines of carbon in the stellar atmospheres, presumably the result of helium burning, and indicating their status as evolved massive stars. WC stars also tend to have lower masses than other WR subtypes; their present-day mass distribution peaks in the 5--15 $M_\odot$ range \citep{Smith2008}. For the LMC population, most (21/23 or $91\pm28$\%) WC stars have a WC4 spectral subtype \citep{Neugent2018}. 

There are a total of 23 WC stars in the LMC \citep{Neugent2018}.  Our Gaia sample with usable proper motion data has 16 stars, with 7 in binaries and 9 single stars. Only 7 out of 23 or $30\pm13$\% of all WC stars are in the 30 Dor region (Table~\ref{tab:subpop}), implying that smaller clusters are also responsible for generating both binary and single WC stars (Figure~\ref{fig:vector_WC}), and thus that WC stars originate from lower-mass progenitors. None of the WC stars with available luminosities have $\log L/L_\odot > 5.9$, our threshold for VMS status.

\begin{figure*}[ht!]
    \centering
    \includegraphics[width=1.0\linewidth]{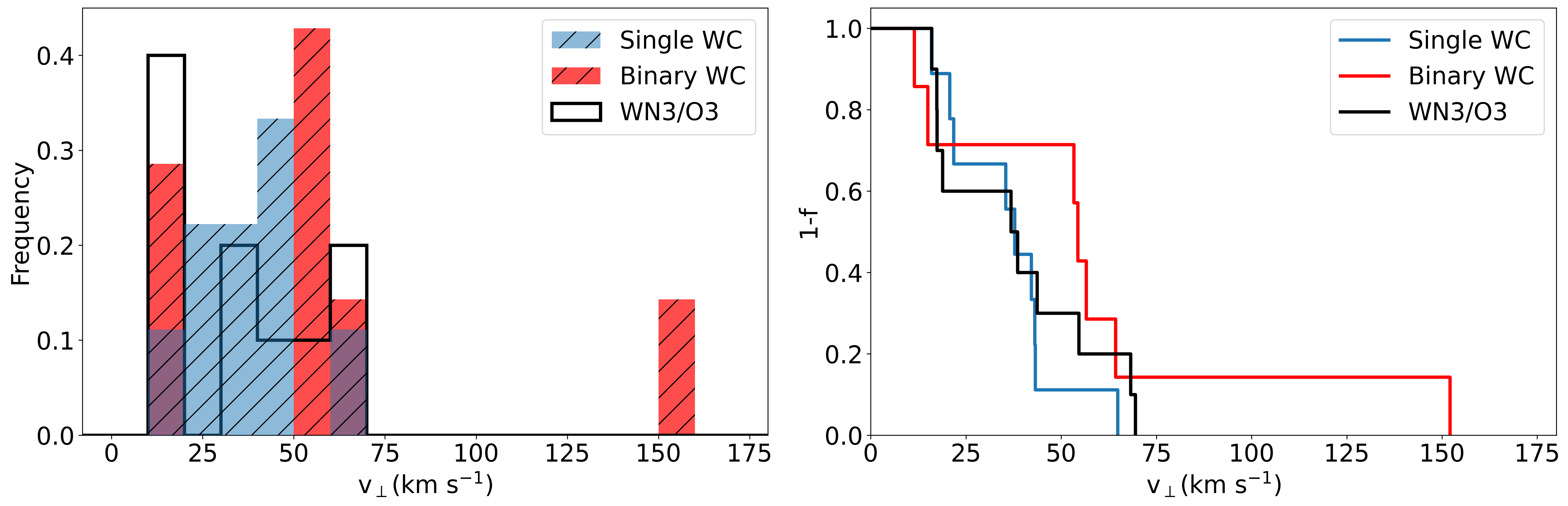}
    \caption{Left panel:  Normalized velocity distribution for single (blue) and binary (red) WC stars. The black outline shows the distribution for WN3/O3 stars.  Right panel: Velocity survival functions for the same groups, color-coded as shown. 
    \label{fig:cdf}}
\end{figure*}

\begin{figure*}
\hspace*{-0.75in}
    \includegraphics[width=1.2\linewidth]{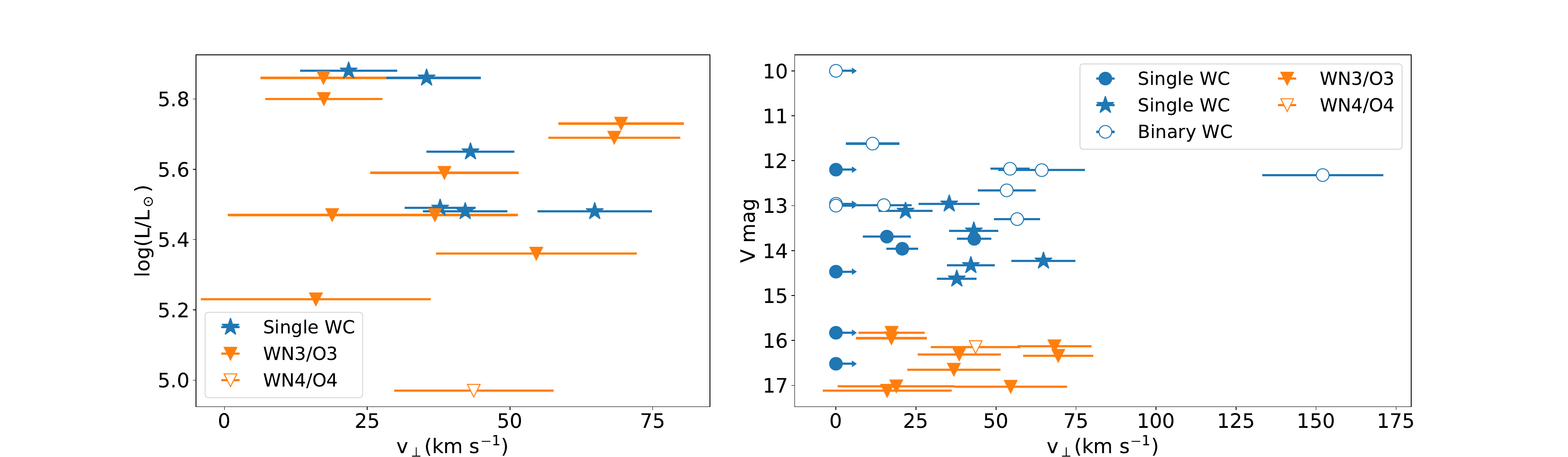}
    \caption{Left panel: Luminosity vs our measured $v_\perp$ for the WC and WN3/O3 stars for which luminosity data is available in the literature from \citet{Aadland2022a, Aadland2022b} and \citet{Neugent2017}, respectively, along with the WN4/O4 star studied by \citet{Massey2024}.  Right panel: $V$ magnitude vs $v_\perp$ for all of the WC stars and WN3/O3 stars in our Gaia samples. The blue stars show the single WC stars that appear in both panels. Arrows indicate lower limits for objects with no Gaia velocity measurements; these objects are likely still bound within clusters (see text). }
    \label{fig:magvsvel}
\end{figure*} 

We find that our Gaia sample WC stars on average have very high peculiar velocities, with median $v_\perp$ of 42 \kms\ (Table~\ref{tab:subpop}). This represents primarily the WC stars ejected from clusters, noting that about 1/4 of all WC stars are dropped from our sample and may correspond to unejected objects.  In particular, of the 7 dropped WC stars, 2 are in the 30 Dor region and 2 more are in crowded regions. Thus, our velocity distribution should more closely resemble those of field star samples. 

According to \citet{Smith2018}, the total WC population is one of the most spatially dispersed of the WR populations, second only to the WN3/O3 stars (Section~\ref{sec:wnodiscussion}). This is consistent with our Gaia WC sample being dominated by walkaways and runaways, and that some have had time to evolve and disperse. The \citet{Hung2021} data may support this scenario, showing that $26\pm12$\% of all WC stars are in Class 3 \hii\ regions (Table~\ref{tab:HII}), compared to 0\% for the VMS types (Section~\ref{sec:VMS}), although we caution that the statistical significance of this comparison is low.

\subsection{Binary WC Stars}

The WC stars have the highest pre-SN binary frequency among LMC WR stars, where 10 out of 23, or $43\pm 16$\%, are binaries (Table~\ref{tab:subpop}). Therefore, it is reasonable to expect that the binary formation channel is likely to be significant for these stars in the lower-metallicity LMC. All of the binary companions in our sample are consistent with OB-star spectral types.

Figure~\ref{fig:cdf} shows that the binary WC stars move faster on average than single WC stars, having a median of 
% $54\pm 22$ \kms\ compared to a median of $38\pm 6.3$ \kms\ 
$54$ \kms\ compared to a median of $38$ \kms\ 
for single stars (Table~\ref{tab:subpop}). A one-sided K-S test returns a $D$-value statistic of $0.60$, corresponding to a statistically significant $p$-value of $0.04$ for the probability that the single WC are faster than the binaries. If confirmed, this would imply that {\it the single WC stars cannot be post-SN survivors of the binary WC systems.} Since, in the conventional binary formation scenario, the WC star is the mass donor and more massive star, it is expected to expire first, leaving the non-WR companion as the survivor. This scenario therefore supports simple binary mass transfer as the origin of the binary WC stars, while the single WC stars likely have a different origin. Figure~\ref{fig:magvsvel} shows that the binaries also have systematically higher luminosities than single WC stars, further supporting the scenario that they represent a fundamentally different population.

Pre-SN binaries can only be ejected through the dynamical mechanism, which generates velocities largely independent of stellar mass \citep{Fujii2011, Banerjee2012}. More massive clusters are expected to generate higher ejection velocities \citep[e.g.,][]{Oh2015}, but no WC binaries are found in R136, and only 1 ejected WC binary appears to originate from that super star cluster (Figure~\ref{fig:vector_WC}).  Thus, the high binary velocities cannot be attributed to the influence of R136, and instead show that lower-mass clusters generate these observed high ejection velocities.  While R136 has ejected many runaways \citep[e.g.,][]{Lennon2018, Renzo2019b, Stoop2024}, they may not yet have evolved into WC stars.  This is consistent with results from simulations by \citet{Farias2019} showing that low-mass clusters can produce more energetic dynamical ejections under slower star-forming conditions.

The WC spatial separation from O stars \citep{Smith2016} suggests that the binaries are ejected prior to the development of WC features, and perhaps prior to binary mass transfer.  This is consistent with expected ejection timescales, which, as noted earlier, can be short, on the order of 1 Myr in dense stellar regions \citep[e.g.,][]{Oh2015, Oh2016, Farias2019} such as those identified with the WC binaries. 

\subsection{Single WC Stars}

In the left panel of Figure \ref{fig:magvsvel}, we see an apparent anticorrelation between luminosity and proper motion velocity for the single WC stars, such that for the ejected population, faster stars are fainter, and presumably less massive. The Kendall rank correlation test evaluates the statistical significance of correlations between two parameters. It gives a $\tau$ statistic of $-0.69$ for luminosity vs velocity of the single WC stars for which luminosity estimates are available \citep{Aadland2022a, Aadland2022b}, corresponding to a $p$-value of 0.055 for the probability that the two quantities are not correlated. This suggests that the trend may be real, although the test includes only the 6 single WC stars in Figure~\ref{fig:magvsvel}, and should be regarded with caution. The right panel of Figure~\ref{fig:magvsvel} shows the observed $V$ magnitude vs $v_\perp$ for all 9 of the single WC stars with Gaia measurements, and we see that qualitatively, the data may still be consistent with a correlation.  We note that all of these stars have the same WC4 spectral type and therefore should have similar bolometric corrections.

If this trend is real, the single WC stars may be secondaries that are survivors of SN ejections, since such a trend is expected for the SN acceleration mechanism \citep[e.g.,][]{Renzo2019a}, and not for dynamical ejections.  For some stars, two-step ejection can occur, in which a binary system first undergoes dynamical ejection then the SN explosion reaccelerates the survivor \citep{PflammAltenburg2010}. Given the high velocities of our WC stars, it may be that this two-step ejection process contributes to the acceleration of a substantial fraction of the WC population.  \citet{Phillips2024} suggest that this could be the case for the OBe population, and they also note that Gaia errors could also still modestly inflate the observed velocity distributions. However, \citet{Wagg2025b} find that SN acceleration is unlikely to significantly enhance dynamical ejections.

If the single WC stars are accelerated at least in part by SN ejections, then they must be the survivors of their progenitor binary systems.  This presents a scenario that is less straightforward, since the stripped star ordinarily is expected to be the more massive, donor star.  One possibility is that the mass gainer quickly gains so much mass that it explodes first \citep{Pols1994}, a rare scenario occurring in close binary systems with, e.g., $\log P \lesssim 0.3$ with 16 $M_\odot$ primaries and initial mass ratios $q \gtrsim 0.7$  \citep[see also][]{Renzo2019a}. Binary population synthesis models by \citet{Eldridge2011} suggest that such objects, which would likely be progenitors of Type Ic supernovae, are indeed highly isolated with relatively high velocities, although quantitatively, their predictions differ from our observations by a factor of $\sim 2$. Alternatively, the mass gainer gains so much mass and luminosity that its enhanced stellar winds cause self-stripping \citep{Dray2005}.  

On the other hand, the single WC stars may be accelerated by dynamical ejections.  Thus, they may have evolved as self-stripped single stars \citep{Shenar2020}.  The lower velocities compared to binary WC stars may be driven by systematic differences in the birth clusters such as cluster mass, compactness, and/or primordial binary parameters \citep[e.g.,][]{Oh2015, Oh2016}. Observed velocities also may be enhanced by dynamical ejections in cases where SN acceleration is important.

Alternatively, single WC stars may represent merger remnants that undergo more extreme, or later, self-stripping as suggested by \citet{Li2024}, or in an explosive process similar to that suggested for the single WNE stars above (Section~\ref{sec:merger}).  This is explored by \citet{Podsiadlowski2010}, who link the process to long gamma-ray bursts and Type Ic supernovae.

\section{WN3/O3 Stars}
\label{sec:wnodiscussion}

WN3/O3 stars are the most recently identified type of WR stars. Reported by \citet{Neugent2017}, their spectra show composite traits of both WN3 and O3 stars, yet their luminosities are too low to be binaries of these types, and they have quite low masses ($6-19\ M_\odot$), similar to those of WC stars. To date, all discovered WN3/O3 stars show no detected binary companions, nor are they known X-ray sources \citep{Massey2023}, with the single exception being the one WN4/O4 star, LMCe055-1 \citep[]{Massey2024}. 

All 10 objects known to be in this class are included in our Gaia sample, corresponding to 9 WN3/O3 stars and 1 WN4/O4 star. Their  median $v_\perp = 39$ \kms, similar to that for our single WC stars (Table~\ref{tab:subpop}), which consists primarily of ejected objects, as noted earlier. Because WN3/O3 stars are fainter and less luminous (Table~\ref{tab:Full}), we caution that their measurement errors are larger: the median value for the error on $v_\perp$ is $13$ \kms\ for this population, compared to $8$ \kms\ for the full WR population. 

\begin{figure}[h]
    \includegraphics[width=1.00\linewidth]{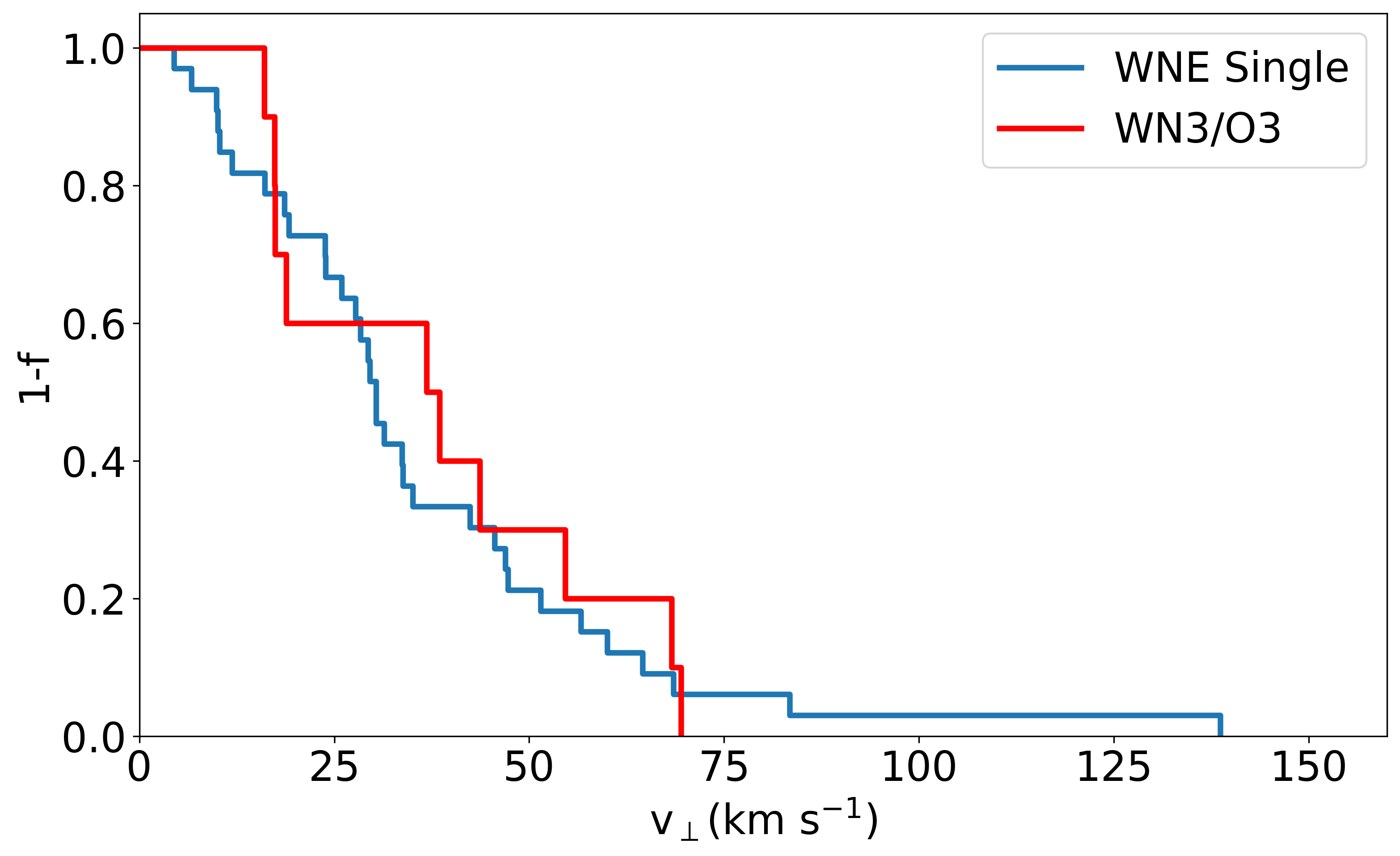}
    \caption{Velocity survival functions for WN3/O3 stars and the remaining LMC WNE stars.}
    \label{fig:WN3O3_WNE_CDF}
\end{figure}

We note that the kinematics of WN3/O3 stars may not be statistically distinct from those of other LMC WNE stars (Figure~\ref{fig:WN3O3_WNE_CDF}).  The two-sided K-S test returns a $p$-value of 0.28 with the $D$ statistic being 0.33, an ambiguous result; the one-sided test gives a $p$-value of 0.14 for the null hypothesis that WNEs move faster than WN3/O3s on average. This may suggest that their origins are not fundamentally different, and formation scenarios for single WNE stars, such as the merger scenario (Section~\ref{sec:merger}), would apply. However, given the possibility of their potentially distinct nature, we consider them as a separate class here.

Using the data of \citet{Hung2021}, $70\pm34$\% of WN3/O3 stars are associated with the most evolved (Class 3) nebular regions, compared to $26\pm12$\% for the WC population (Section~\ref{sec:wcdiscussion}).  Indeed, WN3/O3 stars have the highest percentage of stars in Class 3 \hii\ regions of all the WR subtypes (Table~\ref{tab:HII}). Additionally, \citet{Smith2018} find that these stars are the furthest from their nearest O star neighbors compared to other WR populations.  These data are consistent with the high observed velocities that we obtain.

Their extreme isolation, even compared to WC stars, implies that WN3/O3 stars likely have been traveling for a significantly longer period of time than WC stars. If we consider only our sample of ejected WC stars, assuming that all WC stars omitted from our sample are unejected objects, the median WC separation from O stars is nominally $\sim 80$ pc using the data of \citep{Smith2018}, 
as compared to roughly 160 pc for WN3/O3 stars. This suggests a 2$\times$ longer travel time for WN3/O3 stars. For single WC stars, this is likely a substantial underestimate, given that the Smith et al. analysis includes the binary stars and that we assume the omitted WC stars are all unejected. Thus we conclude that WN3/O3-star travel times are a few times longer than for single WC stars; this may be consistent with longer lifespans if they originate from lower-mass progenitors than WC stars.  As noted above, WN3/O3 stars are mostly associated with the most evolved class of \hii\ regions, and moreso than any other WR subtypes; this can be seen in Figure~\ref{fig:vector_WC} where we see that WN3/O3 are often associated with large superbubble nebulae.  Given the relative isolation of WN3/O3 stars, together with the very short WR lifetimes $\lesssim 0.5$ Myr, they therefore must have been ejected well before they evolved into the WN3/O3 spectral type.

Since WN3/O3 stars seem to be much deeper into the field than WC stars, they are unlikely to be progenitors of the WC stars as suggested by \citet{McClelland2016}. Moreover, Figure \ref{fig:magvsvel} shows that if the correlation between luminosity and velocity for WC stars is real, the WN3/O3 stars do not follow this trend. Thus, the stellar kinematics also do not clearly support a direct evolutionary link between WC and WN3/O3 stars.

In the single-star origin scenarios, the high velocities of WN3/O3 stars must be attributed to dynamical ejection from clusters. Since these stars originate from lower-mass progenitors, they would have longer evolutionary timescales. Therefore, they would have more time to be dynamically ejected from their parent clusters, which could explain their high velocities.

Alternatively, the WN3/O3 stars may be products of binary evolution as proposed by \citet{Smith2018}, in which they are the primary stars that are stripped by their companions, which subsequently explode in a reversal of the conventional supernova sequence \citep{Pols1994}.  This is similar to a scenario suggested above for the origin of single WC stars ejected from clusters, and in principle, it should be rare.  \citet{Smith2018} suggest that for WN3/O3 stars, mass transfer to low-mass secondaries could take place non-conservatively, allowing the mass gainers to remain faint and unexploded. 

% However, the high velocities we find for WN3/O3 stars may suggest that if they have binary origins, they are post-SN, two-step ejections. 
Following modeling by \citet{Gotberg2017}, \citet{Smith2018} find that binary systems with primary initial masses of $12-16\ M_\odot$ can agree with observed characteristics of WN3/O3 stars. Their slow rotational velocities may also support this scenario, as the WN3/O3 progenitor loses angular momentum, along with mass, to its companion.
Although such systems with a reversed SN sequence have low frequency, we note that this likely generates both WN and WC stars:  \citet{Eldridge2011} predict progenitors of both Type IIP and Type Ic supernovae through this mechanism, which could correspond to these respective WR classes.  The velocities are expected to be similar to each other, with the lower-mass SN IIP progenitors traveling about 2$\times$ farther than the SN Ic progenitors.  While quantitative values for the predictions differ somewhat, the correspondence of the qualitative properties is intriguing.

On the other hand, \citet{Smith2018} point out that the SN ejection velocity should be slow due to orbital widening expected in such systems; this may be at odds with the high velocities we obtain for the WN3/O3 stars, unless they are dominated by two-step ejections. Non-conservative mass transfer could also substantially increase the ejection velocities \citep[e.g.,][]{Schurmann2025}. A merger process similar to that suggested to explain single WNE stars (Section~\ref{sec:merger}) might be another possibility to explore.  Further studies using binary population synthesis models \citep[e.g.,][]{Eldridge2011, Schurmann2025} are needed to better understand the likelihood of these scenarios.

\citet{Massey2023} find no evidence of radial velocity variations for the WN3/O3 stars, limiting the existence of possible hidden binary companions to masses $\lesssim 2\ M_\odot$ for periods up to 100 days. 
The only exception is the WN4/O4 star LMCe055-1, which may be an unusual triple or quadruple system \citep{Massey2024}.
If the WN3/O3 stars are categorically single objects, this would be consistent with a possible status as post-SN binary survivors, where a companion could be unbound, or could be a neutron star or a higher-mass black hole with wider ($\sim 1000$-day) orbit such as those consistent with post-SN, bound systems predicted from binary population synthesis models \citep[e.g.,][]{VargasSalazar2025, Dray2005}. Such systems would be difficult to detect. 

A binary origin scenario has been proposed by \citet{Richards2024}, who suggest that WN3/O3 stars may be accreting companions of He stars, which are also difficult to detect and/or have exploded. Thus, SN acceleration may be responsible, or at least a factor, in generating the high WN3/O3 velocities. The supernova kick of the exploding secondary could unbind the system \citep{Pols1994}, ejecting the WN3/O3 star with runaway or walkaway velocity. They could also be two-step ejections that were dynamically ejected from clusters and subsequently reaccelerated to higher speed, although \citet{Wagg2025b} have recently shown that BSS explosions would not add significant kinetic energy to DES ejections. 

Thus, assuming WN3/O3 stars are a physically distinct WR subtype, their high velocities and lack of a trend between luminosity and $v_\perp$ suggest that they are accelerated by either two-step ejections or pure dynamical ejections. On the other hand, the lack of examples with known pre-SN companions is consistent with acceleration by  two-step or even pure SN ejection.  Additionally, other possible scenarios exist to explain the evolutionary status of WN3/O3 stars. For example, they could be merger products or stripped primaries in systems that are now wide binaries with low-mass companions (e.g., \citealt{Peng2022}). The low projected rotational velocities suggest that they may be less likely to be the mass gainers in binary mass-transfer systems, although much depends on the interaction history. In any case, models for their origin must also explain their very high peculiar velocities. Further study is needed to definitively eliminate or confirm the various scenarios. 

\section{Conclusion}

We use data from Gaia DR3 to measure the peculiar transverse velocities for 121 of the 156 LMC WR stars in the catalog of \citet{Neugent2018}.  We examine the kinematics of individual subtypes, in particular, WNh, O If*/WN, WNL, WNE, WC, and WN3/O3 stars. 

For WN subtypes with atmospheric H (WNh, O If*/WN, and WNL), the stellar kinematics and spatial distribution support a luminosity threshold of $\log L/L_\odot \gtrsim 5.9$ to discriminate between stars with likely VMS origins and classical WR stars. Almost all of the VMS stars with these subtypes are found within the 30 Dor region, while most of their classical counterparts are found outside of it, supporting the expectation that the VMS stars originate from the R136.  We also find evidence that the peculiar velocity distribution for the VMS stars is bimodal, with up to half of the stars still unejected, while the remainder are dominated by runaway velocities ($v_\perp > 24$ \kms).  Since R136 is too young to generate significant SN ejections, the data imply that the VMS ages are similar to the dynamical ejection timescale, both of which are predicted to be on the order of 1.5 Myr.  

The multi-peaked structure of the combined VMS WNh, O If*/WN, and WNL velocity distribution appears to be similar in character to that of the SMC field OB stars \citep{Phillips2024}, supporting previous suggestions that normal field OB stars are dominated by dynamical ejections \citep{Moe2025, Phillips2024, DorigoJones2020}.  However, the VMS WR stars lack a substantial component of walkaway stars that are seen among the SMC field OB stars.

Outside of 30 Dor, the lower-luminosity, classical contingent of WNh, O If*/WN, and WNL stars are generally consistent with being lower-mass objects originating from lower-mass clusters. Their velocity distribution is also structured and multi-peaked, suggesting dynamical ejections or multiple mechanisms to accelerate this population.  As noted above, there is also a substantial walkaway component with $v_\perp < 30$ \kms, whose origin is unclear.

WNE stars have a significant pre-SN binary fraction (21/60 or $35\pm9$\%) which suggests the importance of binary processes in their evolution. Only $\sim 13\pm5$\% of WNE stars (8/60) are associated with the 30 Dor region, and they mostly originate in other, lower-mass clusters, likely forming with lower initial stellar masses ($ < 40\ M_{\odot}$). This also applies to the WNE binaries, and thus their high, VMS-like luminosities may be due to mass transfer effects that may cause them to be overluminous for their masses, or to gain VMS masses. The velocity distributions for both the binary and single WNE stars are dominated by only a single peak with monotonically decreasing tail toward high velocity. Single WNE stars travel almost twice as fast (median 30 \kms) as binary WNE stars (median 17 \kms) in general, and
both appear to be ejected populations. The lower luminosities of the single stars compared to binary WNE stars suggest that the binary population may not be the precursors of single WNE stars. 

If single WNE stars have a different origin from the binaries, then a binary mass transfer scenario is unlikely, as it would require a stripped primary to be the survivor of a supernova, an improbable scenario given the large number of single WNE stars. An alternative possibility is that single WNE stars are created from stellar mergers, which are expected to be commonplace. The mass ejected during the merger can produce recoil kicks with velocities $\gtrsim30$ \kms, implying that {\it merger kicks offer another mechanism for accelerating stars to high speeds.} We further speculate that the final envelope is stripped by an explosive merger onto the He-burning shell \citep{Ivanova2002a, Ivanova2003}, resulting in a WN star.

Ejected WC stars have a high median velocity (42 \kms), omitting 7 stars that lack usable Gaia data, 4 of which are from dense regions and likely have low velocities. The spatial distribution of WC stars is consistent with their high speed: the data of \citet{Hung2021} show that most WC stars are in Class 2 \hii\ regions, corresponding to moderately evolved regions, while \citet{Smith2018} find that they are the one of the most isolated populations, further indicating that they are dominated by field stars ejected from their originating clusters.

The high fraction of pre-SN WC binaries ($43 \pm 16$\%) suggests that the binary mass-transfer process is important for this subtype. However, our data show that {\it WC binaries travel faster (median 54 \kms) than single WC stars (median 38 \kms). Thus, generally, single WC stars are not likely descendants of the WC binaries,} and therefore, they presumably originate from different evolutionary paths. The fact that WC stars in binaries have systematically different, higher luminosities than the single ones is consistent with this scenario. The binary WC stars are readily explained as mass donors that will eventually explode before their lower-mass companions in dynamically ejected systems, explaining the lack of single WC stars faster than binaries. The data imply that {\it lower-mass clusters generate fast dynamical ejections,} consistent with some numerical predictions \citep[e.g.,][]{Farias2019}.

Single WC stars may show a trend where fainter, and presumably less massive objects, have higher velocities, which may suggest the action of SN kicks. In this scenario, they could also be binary mass donors, but from systems in which the secondaries rapidly gained mass and exploded first \citep{Pols1994}. Another possible scenario is that these secondaries gain enough mass and luminosity to self-strip, and are the surviving components that experienced black hole kicks \citep{Dray2005}.  On the other hand, the relatively high velocities and shape of the velocity distribution may be due to dynamical ejections, and thus self-stripping, single-star evolution is not completely ruled out.

We find that the WN3/O3 stars have peculiar velocities similar to those of the ejected WC stars, with a  median proper motion velocity of 39 \kms\ for the 10 known LMC objects. These high WN3/O3 velocities are consistent with their reported extreme isolation, since they are even more spatially dispersed than the WC stars \citep{Smith2018}. Their association with the most evolved H II regions is also consistent with those data. While the velocities of WN3/O3 stars are similar to those of WC stars, the two groups do not appear to form a single kinematic population, suggesting that they have distinct origins.  The relative isolation of the WN3/O3 stars suggests that their travel times are several times longer than for WC stars. This may be consistent with WN3/O3 stars having lower-mass progenitors, with correspondingly longer lifespans.  Moreover, it is unlikely that they were ejected with the currently observed spectral type. 

The velocity distribution of WN3/O3 stars may be explained by dynamical ejection from their parent clusters, or they may be two-step ejections.  Or, they may originate from binaries as stripped mass donors and survivors of SN ejections as also suggested for single WC stars, for which the mass gainer exploded first \citep{Pols1994}. Constraints for any companions to the WN3/O3 stars and the possible eccentric, invisible companion of the WN4/O4 star are consistent with this scenario. 

Overall, the velocities of WR stars in the LMC show a dynamic picture revealing kinematic contributions from both dynamical and binary processes. The kinematic profiles of various WR subtypes reflect the complex intersection between the origin of different WR features and ejection mechanisms. While the observed velocities clarify different aspects of both single-star and binary origins, further study is needed to determine the roles of all the different WR subtypes in massive star evolution.
 
\begin{acknowledgments}

We thank Julian Deman and Grant Phillips for providing initial code for calculating the proper motion velocities. We also thank Paul Crowther, Ilya Mandel, Andreas Sander, Tomer Shenar, and other participants of the Aspen Workshop, ``Cosmic Change Agents: Massive Stars in the Early Universe"; this work greatly benefited from discussions with them at the Aspen Center for Physics, which is supported by National Science Foundation grant PHY-2210452.  We also appreciate useful discussions with Juan Farias. Finally, we thank our anonymous referee for helpful feedback.
N.I. acknowledges funding from NSERC under Discovery grant No. RGPIN-2025-05603, and M.R. 
and F. H. acknowledge
support from NASA (ATP: 80NSSC24K0932). This work has made use of data from the European Space Agency (ESA) mission {\it Gaia} (\url{https://www.cosmos.esa.int/gaia}), processed by the {\it Gaia} Data Processing and Analysis Consortium (DPAC, \url{https://www.cosmos.esa.int/web/gaia/dpac/consortium}). Funding for the DPAC has been provided by national institutions, in particular the institutions participating in the {\it Gaia} Multilateral Agreement.
\end{acknowledgments}

\vspace{5mm}
\facilities{Gaia}

\appendix
\vspace*{-0.2in}
\section{WR proper motion velocities}

Table~\ref{tab:Full} presents our measured residual proper motion velocities of our Gaia sample stars relative to their local fields, as described in Section~\ref{sec:meas}.
Columns 1 and 2 give the target star IDs and spectral type from \citet{Neugent2018}. Column 3 is the number of field stars within $3\arcmin$ of the target star. Columns 4 and 5 are the target star velocities in the R.A. and decl. directions. Columns 6 and 7 are the field velocities in R.A. and decl. Column 8 is the residual transverse velocity for each star. All errors shown are measurement errors calculated as by \citet{Phillips2024}. 
Column 9 gives the stellar luminosities from the literature.

\begin{longrotatetable}
\begin{deluxetable*}{ccccccccc}
\label{tab:Full}
\tablecaption{Proper Motion Velocities for WR stars in the LMC}
\tablewidth{0pt}
\tablehead{
\colhead{\parbox[t]{1.5cm}{\centering Star \\ Name(s)}} & 
\colhead{\parbox[t]{1.5cm}{\centering Spectral Type\\ \vphantom{}}} & 
\colhead{\parbox[t]{2cm}{\centering Number of \\ Field Stars}} & 
\colhead{\parbox[t]{2cm}{\centering $v_\alpha$ (target) \\ km s$^{-1}$}} & 
\colhead{\parbox[t]{2cm}{\centering $v_\delta$ (target) \\ km s$^{-1}$}} & 
\colhead{\parbox[t]{1.5cm}{\centering $v_\alpha$ (field) \\ km s$^{-1}$}} & 
\colhead{\parbox[t]{1.5cm}{\centering $v_\delta$ (field) \\ km s$^{-1}$}} & 
\colhead{\parbox[t]{1.5cm}{\centering $v_\perp$ \\ km s$^{-1}$}} & 
\colhead{\parbox[t]{1.5cm}{\centering $\log_{10}{L/L_{\odot}}$\tablenotemark{a}}}
}
\decimalcolnumbers
\startdata
BAT99 1      & WN3             & 66                    & 337.3 $\pm$ 7.4                           & -126.1 $\pm$ 7.0 & 458.5 $\pm$ 4.0 & -58.8 $\pm$ 3.6 & 138.6 $\pm$ 8.3  & 5.30 \\
BAT99 2      & WN2             & 170                   & 477.2 $\pm$ 12.7                          & -48.0 $\pm$ 12.1 & 453.5 $\pm$ 2.6 & -44.6 $\pm$ 2.7 & 23.9 $\pm$ 13.0  & 5.37 \\
BAT99 3      & WN3             & 98                    & 398.5 $\pm$ 4.6                           & -5.7 $\pm$ 5.0   & 404.2 $\pm$ 2.5 & -34.7 $\pm$ 2.4 & 29.5 $\pm$ 5.5   & 5.51 \\
LMC 15666    & WN3+O6V         & 260                   & 459.9 $\pm$ 6.0                           & -26.8 $\pm$ 6.3  & 456.5 $\pm$ 2.1 & -32.0 $\pm$ 1.9 & 6.2 $\pm$ 6.5    & $\cdots$\\
BAT99 5      & WN2             & 282                   & 469.4 $\pm$ 10.1                          & -9.7 $\pm$ 11.7  & 441.7 $\pm$ 2.1 & -22.2 $\pm$ 2.5 & 30.3 $\pm$ 10.6  & 5.45 \\
BAT99 5a     & WN3             & 226                   & 472.6 $\pm$ 6.8                           & -20.3 $\pm$ 8.0  & 443.4 $\pm$ 2.6 & -17.7 $\pm$ 3.0 & 29.3 $\pm$ 7.3   & $\cdots$ \\
BAT99 7      & WN3pec          & 128                   & 407.4 $\pm$ 5.0                           & -86.3 $\pm$ 4.3  & 415.0 $\pm$ 2.9 & -30.2 $\pm$ 3.4 & 56.6 $\pm$ 5.5   & 5.84 \\
BAT99 8      & WC4             & 377                   & 404.9 $\pm$ 8.3                           & 23.1 $\pm$ 11.6  & 454.2 $\pm$ 1.8 & -19.1 $\pm$ 2.0 & 64.8 $\pm$ 10.0  & 5.48 \\
BAT99 9      & WC4             & 92                    & 375.4 $\pm$ 6.7                           & 12.1 $\pm$ 7.0   & 405.1 $\pm$ 3.0 & -17.8 $\pm$ 2.7 & 42.2 $\pm$ 7.4   & 5.48 \\
BAT99 11     & WC4             & 327                   & 384.0 $\pm$ 8.8                           & -11.1 $\pm$ 11.3 & 416.1 $\pm$ 2.2 & -25.9 $\pm$ 2.4 & 35.4 $\pm$ 9.5   & 5.86 \\
LMCe055-1    & WN4/O4          & 247                   & 433.9 $\pm$ 14.3                          & 21.5 $\pm$ 13.6  & 456.1 $\pm$ 2.3 & -16.0 $\pm$ 2.2 & 43.7 $\pm$ 14.0  & $\cdots$\\
BAT99 12     & O2If*/WN5       & 128                   & 355.6 $\pm$ 6.1                           & -56.8 $\pm$ 7.1  & 421.3 $\pm$ 3.0 & -23.4 $\pm$ 3.4 & 73.8 $\pm$ 7.0   & 5.80  \\
BAT99 13     & WN10            & 99                    & 423.7 $\pm$ 4.9                           & -27.0 $\pm$ 5.2  & 404.8 $\pm$ 3.3 & -18.4 $\pm$ 3.5 & 20.8 $\pm$ 6.0   & 5.56 \\
BAT99 15     & WN3             & 168                   & 381.8 $\pm$ 4.4                           & 45.5 $\pm$ 5.5   & 422.0 $\pm$ 2.9 & -9.9 $\pm$ 3.2  & 68.5 $\pm$ 6.0   & 5.57 \\
BAT99 15a    & WN3+abs         & 352                   & 515.2 $\pm$ 7.3                           & -16.2 $\pm$ 8.7  & 445.3 $\pm$ 2.4 & 15.7 $\pm$ 2.6  & 76.8 $\pm$ 7.9   & $\cdots$\\
BAT99 16     & WN7h            & 122                   & 408.7 $\pm$ 7.0                           & 52.1 $\pm$ 7.4   & 399.7 $\pm$ 3.0 & -0.3 $\pm$ 3.0  & 53.1 $\pm$ 7.9   & 5.80  \\
BAT99 17     & WN4             & 235                   & 476.1 $\pm$ 4.5                           & -3.1 $\pm$ 5.9   & 481.9 $\pm$ 2.7 & 5.1 $\pm$ 2.5   & 10.1 $\pm$ 6.1   & 5.69 \\
LMC277-2     & WN3/O3          & 156                   & 414.0 $\pm$ 8.9                           & -3.3 $\pm$ 10.6  & 426.7 $\pm$ 2.9 & 8.6 $\pm$ 3.5   & 17.4 $\pm$ 10.3  & 5.80 \\
BAT99 18     & WN3h            & 240                   & 494.2 $\pm$ 6.5                           & 27.5 $\pm$ 7.3   & 487.3 $\pm$ 2.5 & 6.6 $\pm$ 2.6   & 22.0 $\pm$ 7.6   & 5.63\\
LMC079-1    & WN3/O3          & 281                   & 527.6 $\pm$ 12.2                          & 30.4 $\pm$ 14.6  & 493.8 $\pm$ 2.1 & 12.0 $\pm$ 2.6  & 38.5 $\pm$ 13.0  & 5.59 \\
BAT99 19     & WN3+OB          & 615                   & 416.1 $\pm$ 4.5                           & 26.3 $\pm$ 4.6   & 438.6 $\pm$ 2.3 & 32.1 $\pm$ 2.2  & 23.2 $\pm$ 5.1   & 6.14 \\
BAT99 20     & WC4             & 592                   & 397.1 $\pm$ 4.9                           & 46.6 $\pm$ 4.9   & 438.3 $\pm$ 2.3 & 33.8 $\pm$ 2.2  & 43.2 $\pm$ 5.4   & $\cdots$\\
BAT99 21     & WN4+OB          & 239                   & 383.8 $\pm$ 6.0                           & 11.9 $\pm$ 6.7   & 385.0 $\pm$ 2.2 & 24.8 $\pm$ 2.0  & 12.9 $\pm$ 7.0   & 6.30 \\
BAT99 23     & WN3             & 186                   & 407.2 $\pm$ 13.0                          & 42.3 $\pm$ 14.2  & 387.9 $\pm$ 2.5 & 24.8 $\pm$ 2.2  & 26.0 $\pm$ 13.8  & 5.55 \\
BAT99 24     & WN3             & 621                   & 452.8 $\pm$ 5.4                           & 16.0 $\pm$ 5.7   & 448.1 $\pm$ 2.6 & 50.8 $\pm$ 2.5  & 35.1 $\pm$ 6.3   & 5.54 \\
LMCe132-1    & O3.5If*/WN5     & 190                   & 337.4 $\pm$ 6.4                           & 12.1 $\pm$ 6.5   & 383.6 $\pm$ 2.2 & 26.3 $\pm$ 2.8  & 48.3 $\pm$ 6.8   & $\cdots$\\
BAT99 25     & WN4ha           & 122                   & 546.5 $\pm$ 8.9                           & -8.8 $\pm$ 9.8   & 495.3 $\pm$ 3.0 & 48.1 $\pm$ 3.5  & 76.6 $\pm$ 10.0  & 5.55 \\
BAT99 26     & WN4             & 622                   & 397.8 $\pm$ 5.5                           & 62.6 $\pm$ 6.0   & 449.3 $\pm$ 2.6 & 63.9 $\pm$ 2.7  & 51.5 $\pm$ 6.1   & 5.62  \\
BAT99 27     & BI+WN4          & 706                   & 413.6 $\pm$ 5.9                           & 84.6 $\pm$ 7.5   & 422.7 $\pm$ 2.0 & 70.7 $\pm$ 2.2  & 16.6 $\pm$ 7.4   & 7.30 \\
BAT99 28     & WC6+O5-6        & 768                   & 489.3 $\pm$ 10.2                          & 121.5 $\pm$ 14.6 & 452.0 $\pm$ 2.4 & 69.1 $\pm$ 2.9  & 64.3 $\pm$ 13.5  & $\cdots$\\
LMCe169-1    & WN3/O3          & 98                    & 389.8 $\pm$ 18.4                          & 58.0 $\pm$ 22.3  & 377.0 $\pm$ 2.9 & 67.6 $\pm$ 3.1  & 16.0 $\pm$ 20.2  & 5.23 \\
BAT99 30     & WN6h            & 145                   & 380.4 $\pm$ 4.7                           & 77.3 $\pm$ 4.9   & 377.0 $\pm$ 2.4 & 75.3 $\pm$ 2.5  & 3.9 $\pm$ 5.4    & 5.65  \\
BAT99 31     & WN3             & 248                   & 395.4 $\pm$ 6.3                           & 88.8 $\pm$ 6.7   & 400.3 $\pm$ 2.1 & 65.5 $\pm$ 1.9  & 23.8 $\pm$ 7.0   & 5.33 \\
BAT99 32     & WN6h            & 131                   & 533.7 $\pm$ 8.2                           & 69.4 $\pm$ 8.1   & 491.4 $\pm$ 2.7 & 69.3 $\pm$ 3.5  & 42.3 $\pm$ 8.7   & 5.94  \\
BAT99 33     & Ofpe/WN9        & 150                   & 394.6 $\pm$ 4.1                           & 67.7 $\pm$ 4.2   & 392.3 $\pm$ 2.7 & 70.5 $\pm$ 2.8  & 3.6 $\pm$ 5.0    & 6.50 \\
BAT99 34     & WC4+abs         & 152                   & 543.7 $\pm$ 8.5                           & 74.9 $\pm$ 9.7   & 490.5 $\pm$ 3.0 & 79.1 $\pm$ 3.4  & 53.3 $\pm$ 9.1   & $\cdots$\\
BAT99 35     & WN3             & 72                    & 382.0 $\pm$ 6.6                           & 115.4 $\pm$ 6.5  & 369.6 $\pm$ 4.1 & 70.2 $\pm$ 4.3  & 46.9 $\pm$ 7.8   & 5.60 \\
BAT99 36     & WN3/WCE+OB      & 179                   & 373.1 $\pm$ 6.0                           & 86.7 $\pm$ 5.3   & 384.7 $\pm$ 3.2 & 84.1 $\pm$ 3.3  & 11.9 $\pm$ 6.8   & 5.71 \\
BAT99 37     & WN3             & 132                   & 339.8 $\pm$ 9.1                           & 105.6 $\pm$ 10.0 & 366.4 $\pm$ 2.1 & 95.7 $\pm$ 2.7  & 28.4 $\pm$ 9.4   & 5.65 \\
BAT99 38     & WC4+abs         & 212                   & 376.4 $\pm$ 7.8                           & 99.3 $\pm$ 8.2   & 368.6 $\pm$ 2.3 & 90.9 $\pm$ 2.2  & 11.4 $\pm$ 8.4   & $\cdots$\\
LMCe159-1    & WN3/O3          & 106                   & 341.9 $\pm$ 10.9                          & 160.0 $\pm$ 10.4 & 373.3 $\pm$ 3.1 & 98.1 $\pm$ 3.3  & 69.5 $\pm$ 11.0  & 5.73 \\
BAT99 39     & WC4+O6III/V     & 396                   & 522.0 $\pm$ 19.1                          & 166.7 $\pm$ 16.6 & 381.4 $\pm$ 2.5 & 109.0 $\pm$ 2.3 & 152.1 $\pm$ 18.9 & $\cdots$\\
BAT99 40     & WN4             & 400                   & 323.0 $\pm$ 18.6                          & 165.7 $\pm$ 21.5 & 384.3 $\pm$ 2.5 & 109.2 $\pm$ 2.3 & 83.4 $\pm$ 20.1  & 5.62 \\
BAT99 41     & WN4             & 444                   & 364.5 $\pm$ 5.7                           & 64.1 $\pm$ 5.7   & 397.1 $\pm$ 2.4 & 98.4 $\pm$ 2.5  & 47.3 $\pm$ 6.2   & 5.60 \\
BAT99 42     & B3I+WN5         & 421                   & 371.1 $\pm$ 13.5                          & 79.2 $\pm$ 13.8  & 381.8 $\pm$ 2.3 & 110.6 $\pm$ 2.2 & 33.3 $\pm$ 14.0  & 8.00  \\
BAT99 43     & WN3+OB          & 325                   & 475.4 $\pm$ 5.6                           & 101.0 $\pm$ 6.8  & 474.4 $\pm$ 2.7 & 102.6 $\pm$ 2.9 & 1.8 $\pm$ 7.1    & 5.85 \\
BAT99 44     & WN7h            & 564                   & 388.8 $\pm$ 4.5                           & 82.3 $\pm$ 4.5   & 399.6 $\pm$ 1.9 & 93.7 $\pm$ 2.3  & 15.7 $\pm$ 5.0   & 5.66 \\
BAT99 46     & WN4             & 440                   & 368.0 $\pm$ 7.7                           & 104.5 $\pm$ 7.8  & 399.2 $\pm$ 1.9 & 101.7 $\pm$ 1.9 & 31.4 $\pm$ 7.9   & 5.44 \\
LMC199-1     & WN3/O3          & 563                   & 422.8 $\pm$ 13.8                          & 76.0 $\pm$ 14.9  & 399.4 $\pm$ 1.7 & 104.5 $\pm$ 2.0 & 36.8 $\pm$ 14.6  & 5.47 \\
BAT99 47     & WN3             & 390                   & 387.9 $\pm$ 8.1                           & 120.3 $\pm$ 9.1  & 387.0 $\pm$ 2.1 & 113.7 $\pm$ 1.9 & 6.7 $\pm$ 9.3    & $\cdots$ \\
LMC170-2     & WN3/O3          & 416                   & 376.3 $\pm$ 9.7                           & 53.8 $\pm$ 11.8  & 414.1 $\pm$ 2.7 & 110.6 $\pm$ 2.8 & 68.3 $\pm$ 11.6  & 5.69 \\
BAT99 48     & WN3             & 340                   & 351.3 $\pm$ 10.5                          & 62.1 $\pm$ 11.0  & 392.7 $\pm$ 2.5 & 111.6 $\pm$ 2.9 & 64.6 $\pm$ 11.1  &  5.40\\
BAT99 49     & WN3+O7.5        & 233                   & 484.3 $\pm$ 6.1                           & 102.2 $\pm$ 7.3  & 483.3 $\pm$ 3.1 & 113.3 $\pm$ 3.5 & 11.1 $\pm$ 8.1   & 6.34 \\
BAT99 50     & WN5h            & 426                   & 422.3 $\pm$ 11.8                          & 110.0 $\pm$ 11.6 & 399.8 $\pm$ 2.4 & 110.1 $\pm$ 2.3 & 22.5 $\pm$ 12.1  & 5.65 \\
BAT99 51     & WN3             & 242                   & 398.1 $\pm$ 7.1                           & 125.5 $\pm$ 7.8  & 396.6 $\pm$ 3.1 & 113.7 $\pm$ 2.8 & 11.9 $\pm$ 8.3   & 5.30  \\
BAT99 52     & WC4             & 133                   & 355.7 $\pm$ 6.9                           & 131.9 $\pm$ 7.2  & 389.6 $\pm$ 2.9 & 105.3 $\pm$ 3.4 & 43.1 $\pm$ 7.7   & 5.65 \\
BAT99 54     & WN9             & 346                   & 383.0 $\pm$ 4.8                           & 97.5 $\pm$ 4.6   & 395.6 $\pm$ 2.1 & 116.4 $\pm$ 2.3 & 22.7 $\pm$ 5.2   & 5.75  \\
BAT99 55     & WN11            & 317                   & 438.9 $\pm$ 7.8                           & 88.9 $\pm$ 7.8   & 399.7 $\pm$ 2.7 & 121.8 $\pm$ 2.5 & 51.1 $\pm$ 8.2   & 5.77 \\
BAT99 56     & WN3             & 156                   & 342.2 $\pm$ 5.7                           & 111.9 $\pm$ 6.7  & 374.6 $\pm$ 3.0 & 102.9 $\pm$ 2.9 & 33.7 $\pm$ 6.5   & 5.56 \\
BAT99 57     & WN3             & 229                   & 390.1 $\pm$ 5.5                           & 93.8 $\pm$ 7.3   & 381.7 $\pm$ 1.5 & 110.4 $\pm$ 2.1 & 18.6 $\pm$ 7.2   & 5.40  \\
BAT99 58     & WN7h            & 87                    & 391.1 $\pm$ 5.2                           & 132.1 $\pm$ 6.1  & 383.6 $\pm$ 4.1 & 119.8 $\pm$ 4.1 & 14.4 $\pm$ 7.1   & 5.64 \\
BAT99 59     & WN3+OB          & 129                   & 353.4 $\pm$ 5.6                           & 106.2 $\pm$ 6.0  & 374.8 $\pm$ 2.7 & 114.3 $\pm$ 2.7 & 22.8 $\pm$ 6.3   & 6.45 \\
BAT99 60     & WN3+OB          & 393                   & 353.9 $\pm$ 8.3                           & 108.0 $\pm$ 11.8 & 413.1 $\pm$ 2.6 & 120.7 $\pm$ 2.9 & 60.5 $\pm$ 8.8   & 5.78  \\
BAT99 61     & WC4             & 451                   & 418.0 $\pm$ 8.1                           & 103.1 $\pm$ 8.1  & 423.2 $\pm$ 2.3 & 124.2 $\pm$ 2.5 & 21.7 $\pm$ 8.5   & 5.88 \\
Sk-69 194    & B0I+WN          & 496                   & 441.3 $\pm$ 6.4                           & 110.5 $\pm$ 6.5  & 421.7 $\pm$ 2.2 & 122.6 $\pm$ 2.3 & 23.1 $\pm$ 6.8   & $\cdots$\\
BAT99 62     & WN3             & 147                   & 391.6 $\pm$ 6.4                           & 152.0 $\pm$ 7.0  & 373.3 $\pm$ 2.6 & 127.7 $\pm$ 2.6 & 30.4 $\pm$ 7.2   & 5.41 \\
BAT99 63     & WN4h            & 114                   & 383.9 $\pm$ 6.0                           & 138.5 $\pm$ 7.2  & 376.0 $\pm$ 3.3 & 121.9 $\pm$ 3.7 & 18.3 $\pm$ 7.9   & 5.58 \\
BAT99 64     & WN3+O           & 520                   & 417.8 $\pm$ 5.4                           & 109.8 $\pm$ 5.6  & 418.5 $\pm$ 2.0 & 126.3 $\pm$ 2.2 & 16.6 $\pm$ 6.1   & 6.05 \\
LMC172-1$^*$       & WN3/O3          & 343                   & 419.6 $\pm$ 10.2                          & 136.1 $\pm$ 12.2 & 404.7 $\pm$ 2.5 & 127.4 $\pm$ 2.6 & 17.3 $\pm$ 11.0  & 5.86 \\
LMC143-1     & WN3+O8-9III     & 419                   & 413.9 $\pm$ 6.1                           & 98.1 $\pm$ 6.2   & 423.5 $\pm$ 2.0 & 127.2 $\pm$ 2.5 & 30.6 $\pm$ 6.7   & $\cdots$\\
BAT99 66     & WN3(h)          & 103                   & 404.2 $\pm$ 8.2                           & 97.4 $\pm$ 8.1   & 391.1 $\pm$ 3.1 & 110.0 $\pm$ 3.2 & 18.2 $\pm$ 8.7   & 5.78 \\
BAT99 67$^*$   & WN5h            & 355                   & 386.1 $\pm$ 4.6                           & 135.4 $\pm$ 4.1  & 386.1 $\pm$ 2.2 & 137.2 $\pm$ 1.8 & 1.8 $\pm$ 4.5    & 5.96 \\
BAT99 68$^*$  & O3.5If*/WN7     & 352                   & 394.1 $\pm$ 4.7                           & 134.2 $\pm$ 4.1  & 385.4 $\pm$ 1.9 & 135.9 $\pm$ 1.7 & 8.9 $\pm$ 5.1    & 6.00 \\
BAT99 70$^*$       & WC4             & 343                   & 397.9 $\pm$ 4.9                           & 121.4 $\pm$ 4.5  & 384.8 $\pm$ 2.0 & 137.5 $\pm$ 1.7 & 20.7 $\pm$ 5.0   & $\cdots$\\
BAT99 72     & WN4+abs         & 310                   & 379.6 $\pm$ 9.1                           & 135.0 $\pm$ 8.0  & 380.4 $\pm$ 2.0 & 136.8 $\pm$ 1.7 & 1.9 $\pm$ 8.4    & 5.80 \\
BAT99 73     & WN5             & 336                   & 387.2 $\pm$ 6.1                           & 137.6 $\pm$ 5.6  & 384.0 $\pm$ 1.9 & 134.5 $\pm$ 1.8 & 4.4 $\pm$ 6.2    & 5.72 \\
BAT99 74     & WN3+abs         & 320                   & 405.3 $\pm$ 8.9                           & 183.9 $\pm$ 8.2  & 383.5 $\pm$ 1.8 & 137.1 $\pm$ 1.8 & 51.7 $\pm$ 8.5   & 5.69 \\
BAT99 75     & WN4             & 178                   & 411.3 $\pm$ 6.4                           & 81.3 $\pm$ 6.2   & 381.9 $\pm$ 2.9 & 116.1 $\pm$ 2.1 & 45.5 $\pm$ 6.7   & 5.56 \\
BAT99 76     & WN9             & 290                   & 376.0 $\pm$ 4.4                           & 185.6 $\pm$ 4.0  & 380.3 $\pm$ 2.1 & 136.1 $\pm$ 1.8 & 49.7 $\pm$ 4.4   & 5.66 \\
BAT99 77$^*$       & WN7             & 416                   & 350.6 $\pm$ 7.1                           & 116.8 $\pm$ 6.0  & 386.3 $\pm$ 1.9 & 135.8 $\pm$ 1.6 & 40.4 $\pm$ 7.2   & 6.79 \\
BAT99 79$^*$       & WN7             & 407                   & 394.0 $\pm$ 4.0                           & 139.0 $\pm$ 3.6  & 385.3 $\pm$ 1.8 & 135.4 $\pm$ 1.6 & 9.4 $\pm$ 4.4    & 6.17 \\
BAT99 81     & WN5h            & 102                   & 411.1 $\pm$ 6.6                           & 81.3 $\pm$ 7.0   & 377.6 $\pm$ 3.4 & 114.6 $\pm$ 3.6 & 47.2 $\pm$ 7.6   & 5.48 \\
BAT99 82$^*$       & WN3             & 309                   & 397.8 $\pm$ 11.4                          & 141.9 $\pm$ 11.5 & 389.5 $\pm$ 2.1 & 136.5 $\pm$ 1.9 & 9.9 $\pm$ 11.6   & 5.53\\
BAT99 84     & WC4(+OB)        & 556                   & 413.1 $\pm$ 8.4                           & 132.8 $\pm$ 8.5  & 416.4 $\pm$ 1.6 & 118.1 $\pm$ 1.6 & 15.0 $\pm$ 8.6   & $\cdots$\\
BAT99 85$^*$       & WC4(+OB)        & 250                   & 441.6 $\pm$ 5.6                           & 155.3 $\pm$ 5.4  & 389.7 $\pm$ 2.7 & 139.0 $\pm$ 2.0 & 54.4 $\pm$ 6.2   & $\cdots$\\
BAT99 86$^*$      & WN3             & 222                   & 383.8 $\pm$ 11.4                          & 153.0 $\pm$ 10.4 & 389.7 $\pm$ 2.3 & 138.0 $\pm$ 2.3 & 16.1 $\pm$ 10.8  & 5.33  \\
BAT99 87     & WC4             & 509                   & 422.8 $\pm$ 7.5                           & 111.6 $\pm$ 7.1  & 411.5 $\pm$ 1.9 & 122.8 $\pm$ 1.6 & 15.9 $\pm$ 7.5   & $\cdots$\\
BAT99 88$^*$       & WN3/WCE         & 293                   & 364.2 $\pm$ 17.0                          & 153.2 $\pm$ 14.3 & 390.3 $\pm$ 2.0 & 143.8 $\pm$ 1.9 & 27.7 $\pm$ 16.9  & 5.80 \\
BAT99 89$^*$       & WN6             & 273                   & 387.9 $\pm$ 4.3                           & 159.6 $\pm$ 4.3  & 390.2 $\pm$ 2.0 & 144.0 $\pm$ 1.8 & 15.8 $\pm$ 4.7   & 5.78 \\
BAT99 90$^*$       & WC4             & 205                   & 403.7 $\pm$ 6.0                           & 101.2 $\pm$ 5.5  & 397.9 $\pm$ 2.7 & 138.5 $\pm$ 2.8 & 37.8 $\pm$ 6.2   & 5.49 \\
LMC173-1     & WN3+O7V         & 506                   & 387.2 $\pm$ 8.0                           & 126.9 $\pm$ 6.4  & 412.1 $\pm$ 2.0 & 125.8 $\pm$ 1.6 & 25.0 $\pm$ 8.3   & $\cdots$\\
BAT99 92$^*$       & B1I+WN3         & 203                   & 401.1 $\pm$ 5.3                           & 154.4 $\pm$ 5.1  & 390.9 $\pm$ 2.7 & 143.2 $\pm$ 2.3 & 15.1 $\pm$ 5.8   & 6.95 \\
BAT99 93$^*$       & O3If*           & 262                   & 371.8 $\pm$ 4.2                           & 145.2 $\pm$ 4.2  & 391.7 $\pm$ 2.1 & 147.0 $\pm$ 2.0 & 20.0 $\pm$ 4.7   & 5.90 \\
LMCe063-1$^*$      & WN11            & 274                   & 435.2 $\pm$ 4.7                           & 125.3 $\pm$ 4.6  & 414.1 $\pm$ 2.2 & 125.3 $\pm$ 2.5 & 21.0 $\pm$ 5.2   & $\cdots$\\
BAT99 94$^*$       & WN3/4pec        & 274                   & 407.3 $\pm$ 4.5                           & 118.0 $\pm$ 4.4  & 414.0 $\pm$ 2.2 & 125.8 $\pm$ 2.5 & 10.3 $\pm$ 5.0   & 5.80 \\
BAT99 96$^*$       & WN7             & 572                   & 377.6 $\pm$ 4.4                           & 144.6 $\pm$ 3.9  & 398.0 $\pm$ 2.0 & 150.2 $\pm$ 1.8 & 21.2 $\pm$ 4.8   & 6.35 \\
BAT99 97$^*$       & O3.5If*/WN7     & 559                   & 396.7 $\pm$ 4.3                           & 152.2 $\pm$ 4.0  & 398.2 $\pm$ 1.9 & 149.9 $\pm$ 1.8 & 2.8 $\pm$ 4.5    & 6.30 \\
BAT99 99$^*$       & O2.5If*/WN6     & 602                   & 402.7 $\pm$ 14.5                          & 96.0 $\pm$ 15.6  & 397.6 $\pm$ 1.8 & 151.6 $\pm$ 1.8 & 55.8 $\pm$ 15.7  & 5.90 \\
BAT99 100$^*$      & WN6h            & 615                   & 399.3 $\pm$ 17.4                          & 108.9 $\pm$ 19.5 & 397.4 $\pm$ 1.9 & 151.4 $\pm$ 1.8 & 42.6 $\pm$ 19.6  & 6.15 \\
R136-015$^*$       & O2If*/WN5       & 579                   & 502.9 $\pm$ 20.4                          & 154.5 $\pm$ 20.0 & 397.7 $\pm$ 1.9 & 151.7 $\pm$ 1.8 & 105.3 $\pm$ 20.5 & $\cdots$\\
BAT99 113$^*$      & O2If*/WN5       & 600                   & 411.1 $\pm$ 10.2                          & 144.3 $\pm$ 10.2 & 397.6 $\pm$ 1.9 & 151.6 $\pm$ 1.8 & 15.3 $\pm$ 10.4  & 6.09  \\
BAT99 114$^*$      & O2If*/WN5       & 567                   & 381.1 $\pm$ 8.8                           & 126.0 $\pm$ 8.2  & 397.8 $\pm$ 1.9 & 151.7 $\pm$ 1.9 & 30.7 $\pm$ 8.6   & 6.44 \\
BAT99 116$^*$      & WN4.5h          & 594                   & 434.4 $\pm$ 6.5                           & 115.2 $\pm$ 6.5  & 397.6 $\pm$ 1.9 & 151.6 $\pm$ 1.8 & 51.7 $\pm$ 6.8   & 7.05 \\
BAT99 117$^*$      & WN4.5           & 167                   & 393.1 $\pm$ 6.3                           & 183.7 $\pm$ 8.7  & 399.6 $\pm$ 3.1 & 150.5 $\pm$ 3.4 & 33.8 $\pm$ 9.3   & 6.40 \\
BAT99 118$^*$      & WN5/6+WN6/7   & 209                   & 425.5 $\pm$ 6.4                           & 217.7 $\pm$ 7.4  & 397.8 $\pm$ 2.5 & 156.5 $\pm$ 2.9 & 67.1 $\pm$ 7.8   & 6.66 \\
VFTS682$^*$        & WN5h            & 519                   & 403.2 $\pm$ 9.7                           & 167.8 $\pm$ 11.1 & 398.3 $\pm$ 2.3 & 154.0 $\pm$ 2.1 & 14.7 $\pm$ 11.1  & $\cdots$ \\
BAT99 119$^*$      & WN6+O3.5If*/WN7 & 535                   & 419.9 $\pm$ 5.9                           & 181.1 $\pm$ 6.1  & 397.7 $\pm$ 2.2 & 152.0 $\pm$ 1.9 & 36.6 $\pm$ 6.4   & 6.57 \\
BAT99 120          & WN9             & 337                   & 434.4 $\pm$ 6.5                           & 154.2 $\pm$ 7.3  & 419.2 $\pm$ 1.7 & 119.3 $\pm$ 2.0 & 38.1 $\pm$ 7.4   & 5.58 \\
BAT99 122$^*$      & WN5h            & 152                   & 447.0 $\pm$ 5.3                           & 203.2 $\pm$ 6.4  & 400.4 $\pm$ 2.7 & 150.5 $\pm$ 3.4 & 70.4 $\pm$ 6.7   & 6.23 \\
BAT99 123          & WO3             & 119                   & 404.0 $\pm$ 8.6                           & 205.9 $\pm$ 10.8 & 392.2 $\pm$ 3.3 & 150.6 $\pm$ 3.9 & 56.6 $\pm$ 11.4  & $\cdots$ \\
LMC174-1     & WN3/O3          & 382                   & 427.4 $\pm$ 15.1                          & 128.9 $\pm$ 18.2 & 426.1 $\pm$ 1.6 & 147.7 $\pm$ 1.9 & 18.8 $\pm$ 18.3  & 5.47 \\
BAT99 127    & WC4+O           & 454                   & 471.2 $\pm$ 7.0                           & 135.4 $\pm$ 7.7  & 415.2 $\pm$ 1.7 & 143.3 $\pm$ 1.9 & 56.6 $\pm$ 7.2   & $\cdots$ \\
LMCe078-3    & WN3/O3          & 176                   & 366.6 $\pm$ 17.1                          & 197.0 $\pm$ 17.5 & 403.4 $\pm$ 3.1 & 156.6 $\pm$ 3.1 & 54.6 $\pm$ 17.6  & 5.36 \\
BAT99 129    & WN3+O5V         & 222                   & 467.8 $\pm$ 5.4                           & 157.0 $\pm$ 5.9  & 444.8 $\pm$ 3.0 & 153.3 $\pm$ 2.7 & 23.3 $\pm$ 6.2   & 6.20  \\
BAT99 130    & WN11            & 127                   & 497.4 $\pm$ 5.4                           & 138.5 $\pm$ 5.4  & 395.6 $\pm$ 3.7 & 171.8 $\pm$ 3.3 & 107.1 $\pm$ 6.6  & 5.68 \\
BAT99 131    & WN4             & 197                   & 346.9 $\pm$ 5.6                           & 164.9 $\pm$ 5.7  & 363.6 $\pm$ 2.2 & 155.5 $\pm$ 2.3 & 19.2 $\pm$ 6.1   & 5.67 \\
BAT99 132    & WN4             & 150                   & 372.7 $\pm$ 5.7                           & 191.6 $\pm$ 5.1  & 365.6 $\pm$ 2.9 & 149.8 $\pm$ 2.9 & 42.4 $\pm$ 5.9   & 5.58 \\
BAT99 133    & WN11            & 134                   & $452.6 \pm 6.9$                           & $164.6 \pm 7.0$  & $369.7 \pm 2.6$ & $150.7 \pm 3.0$ & $84.1 \pm 7.4$   & 5.69  \\
BAT99 134    & WN3             & 150                   & $406.8 \pm 6.6$                           & $205.2 \pm 5.8$  & $377.0 \pm 2.6$ & $153.1 \pm 2.8$ & $60.0 \pm 6.6$   & 5.51 \\
\enddata
\tablenotetext{a}{Luminosities for WN, WNh, and O If*/WN stars are taken from \citet{Hainich2014}, luminosities for WC stars are taken from \citet{Aadland2022a,Aadland2022b}, and luminosities for WN3/O3 stars are taken from \citet{Neugent2017}}
\tablenotetext{*}{Objects marked with an asterisk are considered to be within the 30 Dor region (Figure~\ref{fig:vectors_30Dor}), following \citet{Hung2021}.}

\end{deluxetable*}
\end{longrotatetable}
   
\bibliography{Citations}
\bibliographystyle{aasjournal}

\end{CJK*}
\end{document}